\documentclass{aa}  

\usepackage{graphicx}
\usepackage{graphicx}
\usepackage{multicol}
\usepackage{graphicx}
\usepackage[svgnames]{xcolor}
\usepackage{enumitem}
\usepackage{appendix}
\usepackage{comment}

\usepackage[colorlinks=true, linkcolor=DarkBlue, urlcolor = blue, citecolor = DarkBlue,pdfborder={0 0 0}]{hyperref}
\usepackage{lipsum}
\usepackage[normalem]{ulem}

\newcommand{\new}[1]{#1}
\newcommand{\rev}[1]{#1}

\newcommand{\fink}{{\sc Fink}}

\begin{document} 

   \title{Early Identification of Optical Tidal Disruption Events}

   \subtitle{A science module for the Fink broker}

   \author{M. Llamas Lanza$^{\star}$
          \inst{1}
          \and
          S. Karpov\thanks{These authors contributed equally to this work and share first authorship.}
          \inst{2}
          \and
          E. Russeil
          \inst{3}
          \and
          E. Quintin
          \inst{4}
          \and
          E. E. O. Ishida
          \inst{5}
          \and
          J. Peloton
          \inst{6}
          \and
          M.~V.~Pruzhinskaya
          \inst{7}
          \and
          A. Möller
          \inst{8,9}
          }

   \institute{
              IRAP, Université de Toulouse, CNRS, CNES, UPS, Toulouse, France.
              \email{miguel.llamas.lanza@gmail.com}
         \and
             Institute of Physics of the Czech Academy of Sciences, Na Slovance 1999/2, 182 00 Prague 8, Czech Republic
             \email{karpov@fzu.cz}
         \and
             The Oskar Klein Centre, Department of Astronomy, Stockholm University, AlbaNova, SE-10691 Stockholm, Sweden
         \and
             European Space Agency (ESA), European Space Astronomy Centre (ESAC), Camino Bajo del Castillo s/n, 28692 Villanueva de la Cañada, Madrid, Spain 
         \and
             Université Clermont Auvergne, CNRS, LPCA, Clermont-Ferrand, F-63000, France
         \and
             Université Paris-Saclay, CNRS/IN2P3, IJCLab, Orsay, France
         \and
            Lomonosov Moscow State University, Sternberg Astronomical Institute, Universitetsky 13, Moscow 119234, Russia
         \and
             Centre for Astrophysics and Supercomputing, Swinburne University of Technology, Mail Number H29, PO Box 218, 31122 Hawthorn, VIC, Australia
        \and
        ARC Centre of Excellence for Gravitational Wave Discovery (OzGrav), John St, Hawthorn, VIC 3122, Australia
         }
             
   \date{}

 
  \abstract
   {The detection of tidal disruption events (TDEs) is one of the key science goals of large optical time-domain surveys such as the Zwicky Transient Facility (ZTF) and the upcoming Vera C. Rubin Observatory Legacy Survey of Space and Time. However, identifying TDEs in the vast alert streams produced by these surveys requires automated and reliable classification pipelines that can select promising candidates in real time.}
   {We developed a module within the \fink\ alert broker to identify TDEs during their rising phase. It was built to autonomously operate within the ZTF alert stream, producing a list of candidates every night and enabling spectral and multi-wavelength follow-up near peak brightness.}
   {All rising alerts are submitted to selection cuts and feature extraction using the \textsc{Rainbow} multi-band lightcurve fit. Best-fit values were used as input to train an XGBoost classifier with the goal of identifying TDEs. The training set was constructed using ZTF observations for objects with available classification in the Transient Name Server. Finally, candidates with high enough probability were visually inspected.}
   {The classifier achieves 76\% recall, indicating strong performance in early-phase identification, despite the limited available information before the peak.
   We show that, out of the known TDEs that pass the selection cuts, half of them are flagged as TDE before halfway in their rise, proving the feasibility of early classification.
   Additionally, new candidates were identified by applying the classifier on archival data, including a likely repeated TDE and some potential TDEs occurring in active galaxies.
   The module is implemented into the \fink\ alert processing framework, reporting each night a small number of candidates to dedicated communication channels through a user-friendly interface for manual vetting and potential follow-up.
   }
   {}


   \authorrunning{M. Llamas Lanza \& S. Karpov et al.}
   \maketitle
%

\section{Introduction}

Tidal Disruption Events \citep[TDEs, e.g.][]{gezari_tidal_2021} correspond to the destruction of a star passing within the tidal radius of a supermassive black hole, as the intense tidal forces overcome the star's self gravity. Depending on the properties of the star and the black hole, and on their relative trajectories, this disruption can be complete or only partial, in which case it can even be repeated. The typical observational behavior of TDEs is a sudden rise in the emission of the galactic nucleus, well described by a thermal continuum (in the case of non-jetted TDEs), followed by a decay over a few months to a few years, consistent more or less with a $t^{-5/3}$ power-law decay \citep[e.g.][]{1988Natur.333..523R} -- this general behavior is seen across wavelengths, although with different timing and temperature properties.

These events are relatively rare, with a rate of about $10^{-5}$ galaxy$^{-1}$ year$^{-1}$ \citep[e.g.][]{yao_ztfdemographics_2023}. Historically, the first known TDEs were discovered in soft X-rays, by the \textit{ROSAT} satellite \citep[e.g][]{bade_detection_1996}. However, most of the current sample of TDEs was provided by optical surveys, such as the Zwicky Transient Facility \citep[ZTF,][]{bellm_zwicky_2014}, as they perform an intense monitoring of the sky with cadence optimised for finding such transient events. TDEs have also been discovered in radio \citep[e.g.][]{bloom_possible_2011} and infrared \citep[e.g.][]{masterson_infrared_2024}. The current sample contains about 150 TDEs across all wavelengths \citep[e.g.][]{gezari_tidal_2021, langis_tdecatalog_2025}.

While the general picture of a bright, months-long transient event is common to most wavelengths, the precise spectral and timing properties do not always match between them. A lot of questions remain unanswered about the multi-wavelength counterparts of these events \citep[e.g.][]{saxton2018spectral} and about the precise emission mechanisms at work. One current hypothesis is that a central X-ray engine powers all TDEs \citep[e.g.][]{dai2018unified}, although X-ray emission is not always seen in these objects \citep[e.g.][]{guolo_xraytdes}. 

To answer these questions and improve our understanding of the mechanisms at play in TDEs, there are mainly two different possible avenues of progress. The first one is simulations \citep[e.g.][]{Guillochon_Hydrodynamical_2013}; this is particularly difficult for TDEs because of the influence of general relativity, the prevalence of both shocks and plasma effects in the hot ionized matter around the black hole, and the wide range of timescales and spatial scales needed to capture the entirety of the phenomenon. The other possibility is observational. It is possible to constrain the spectral-timing properties of these objects through systematic, and rapid, multi-wavelength follow-up and by monitoring newly detected TDEs. Studying these properties could allow us to constrain the current existing models, e.g. by enabling population studies.

These observational approaches are particularly relevant now, with the advent of large scale transient-focused missions such as the aforementioned ZTF, the recently launched Einstein Probe \citep{einsteinProbe_2015}, or the upcoming Vera C. Rubin Observatory Legacy Survey of Space and Time \citep[LSST;][]{ivezic2019lsst}. These missions produce an unprecedented amount of data, and dedicated software and methods must be developed in order to ensure the identification of TDEs (or any other type of specific transients) in their data streams. \rev{TDE photometric candidate samples can be contaminated by various types of transients, including bright supernovae or superluminous supernovae, as well as high accretion episodes in Active Galactic Nuclei}. Ideally, the combination of spectral and photometric data is necessary to provide a definitive answer as to the nature of a transient event. However, spectral data is observationally expensive, and as such, the community has dedicated considerable effort to the development of photometric classification techniques.

Recently, several classifiers have been proposed to tackle the specific issue of optical TDE photometric classification. They proceed using intrinsically different methods. FLEET \citep{Gomez2023} relies on automatic host galaxy association and lightcurve feature extraction to train a Random Forest classifier. It proposes two separate models, able to classify early and full lightcurves. 
ALeRCE \citep{alerce_tde} proposed to expand their previous multiclass classifier \citep{alerce_previous} with 24 additional features, designed to separate TDEs from other transients. They include distance from the host, new fitting models, and color information. 
NEEDLE \citep{needle} directly leverages the detection and reference images that are combined with photometric information from the alert packets to train a convolutional neural network. It comes in several versions, with or without usage of the host galaxy information. Finally, \texttt{tdescore} \citep{Stein_tdescore_2023} extracts features by performing a gaussian process assuming a linear evolution of the transient color. They are used in complement with host galaxy information to train an XGBoost classifier \citep{xgboost}. 

Despite the many recent efforts on TDE classification pipelines, only FLEET attempts to identify TDEs in their early phase. It proceeds by fitting a parametric model to lightcurves with less than 20 days history and fixing the decay parameter such that the fit effectively concerns only the rising part. Such method offers rapid classification with fewer data points, however it doesn't guarantee that the identification occurs before the peak luminosity of the source. 

In this paper, we focus on the identification of TDEs solely based on the rising part of their lightcurves. Our goal is to provide a reliable pre-maximum classifier that can be used to separate a stream of promising candidates, to enable early follow-up. The classifier should not exclusively fine-tune its predictions based on the current sparse number of known TDEs, but should also remain compatible with our understanding of their possible characteristics. Hence, our aim is not to maximize purity, but rather completeness, such that the classifier can be used as a\rev{n early warning} tool to obtain as many promising candidates as possible directly from the alert stream.
In particular, we developed a science module for the \fink\ broker which will process the alert data stream after every night and report a list of candidates.

\rev{We introduce the \fink\ alert broker in Section 2, which processes ZTF and Rubin LSST data. We then describe the design choices of our TDE early classification module in Section 3. We present the results in Section 4, including the efficiency of the classifier and a selection of candidates it found in the ZTF archive. Finally, we discuss this work in Section 5, putting it into perspective with other classifiers and the upcoming Rubin data.}

\section{The \fink\ alert broker}

\fink\ is an astronomy community broker which processes time-domain alert streams and redirects relevant sub-sets of the data to science teams and follow-up facilities, while also making the data accessible through its web-portal\footnote{\url{https://fink-portal.org/}} and data transfer service\footnote{\url{https://fink-portal.org/download}} \citep{moller2021fink}. The broker collects, filters, aggregates, enriches, and redistributes incoming data streams in real-time. This is done through a series of filters and science modules developed by the community. Independent science teams are responsible for developing modules according to their specific scientific interests. Once finalized, such modules are integrated to the broker and real time processing is centralized in dedicated infrastructures.
\fink\ was conceived to process the full stream of alerts from the Vera C. Rubin Observatory \citep{ivezic2019lsst} as part of the Rubin Community brokers\footnote{Other brokers dedicated to Rubin processing are ALeRCE \citep{Forster:2020}, AMPEL \citep{ampel}, Antares \citep{antares}, Babamul, Lasair \citep{lasair} and Pitt-Google.}.

Since late 2019, \fink\ has been operating on the ZTF public stream \citep{bellm2019ztf}, as a precursor experiment for Rubin. The broker receives data resulting from the difference imaging analysis pipeline, which includes 3 image stamps, contextual information and 30 days photometric history within each alert package.
Alerts that satisfy a set of quality criteria\footnote{The real-bogus score \citep{mahabal2019machine} must be above 0.55, and there should be no prior-tagged bad pixels in a 5~x~5 pixel stamp around the alert position} are then processed by \fink\ science modules, whose task is to add value and thus, help decipher the nature of the sources. These include cross-match with major catalogs as well as statistics and machine learning based routines which transfer domain knowledge to each alert\footnote{\url{https://fink-broker.readthedocs.io/en/latest/broker/science_modules/}}. 
The range of science modules, and the respective astrophysical subjects they tackle, reflects the variety present within the \fink\ community. This includes supernovae \citep[SNe,][]{leoni2022fink, moller2020supernnova, moller2025}, kilonovae \citep{biswas2023enabling}, gamma-ray bursts \citep{masson2024}, microlensing \citep{ban2025}, solar system objects \citep{lemontagner2023}, as well as multi-class classifiers \citep{fraga2024}, extragalactic hostless transients \citep{pessi2024}, and anomaly detection\footnote{\url{https://github.com/fink-anomaly/}}, to cite a few. Each science team has complete autonomy while developing their module. Nevertheless, the output of all \rev{modules} is public and, as such, available to anyone using \fink. Thus, it is crucial to clarify the hypothesis used in the development of each module, so the community can make informed decisions regarding their output (e.g. class probabilities or candidate lists). The broker currently processes around 200 000 ZTF alerts per night, a number which is expected to scale to 10 million once LSST becomes operational. 

The goal of this work is to give details on the development of the Early TDE science module available in \fink. We describe in detail the target candidates and the hypothesis used in its development, thus enabling others to either make informed follow-up decisions based on our list of candidates, or inspire the development of alternative pipelines.

\section{\rev{Early TDE} module design}
\label{sec:module_design}

We developed a supervised classifier aimed at identifying rising TDEs in the ZTF alert stream. Below, we give details on the main steps leading to the final pipeline. These include: selection cuts and identification of rising lightcurves (Section \ref{subsec:prepro}), feature extraction (Section \ref{sec:features}), \rev{filtering and labeling (Section \ref{sec:filtering_labeling})} and the classifier itself (Section \ref{sec:classifier}).

\subsection{Preprocesing}
\label{subsec:prepro}

Starting from ZTF lightcurves available in \fink, we applied a set of filters to select relevant objects. First, we used a positional crossmatch with SIMBAD \citep{wenger2000simbad} and selected objects that are either not present in the catalog or are classified into categories broadly consistent with TDE host environments (e.g., extragalactic, lensing systems, transients, or active galaxies). \rev{Second}, we excluded all known solar system objects present in the Minor Planet Center\footnote{\url{https://www.minorplanetcenter.net/iau/mpc.html}}. \rev{Third, i}n order to limit the contamination from galactic objects, we also excluded all sources with galactic latitude $|b|<20^{\circ}$.
For every remaining object, we extracted from the \fink\ database the complete ZTF lightcurves in the $g$ and $r$ bands covering the time interval between November 2019 and February 2025.

We corrected the photometry for these objects for galactic extinction using the \citet{sfd_map} estimates of $E(B-V)$ and \citet{dustmaps} implementation of the \citet{g23_extinction} extinction law.
We then converted the photometric measurements to fluxes using a magnitude zeropoint of 27.5, which gives the flux in so-called \texttt{FLUXCAL} units \citep{snana}.
Flux values correspond to the brightness differences between the science and reference stamps, with positive meaning that the transient in the science image is brighter than in the reference, and negative marking a fainter object in the science stamp.

For each remaining lightcurve, we performed a sliding window search to select all possible rising segments. To achieve this, at each data point we defined three time intervals (see Figure~\ref{fig:rainbow_example}): a 100-day \textit{fitting} window immediately preceding the data point, used to search for the rising TDE; a 100-day \textit{historical} interval prior to the fitting window; and a \textit{pre-historical} window including all data before that. Then, the following criteria, which must be fulfilled simultaneously in each of these windows, was applied:
\begin{enumerate}[label=\roman*.]
    \item \rev{Fitting window:} 
    \begin{itemize}[label=\textbullet] 
        \item \emph{Rising in at least one band}: Either the last detection in the band must be at least $2\sigma$ above the minimum \rev{flux in this band} or the linear regression slope across the entire time interval should be significantly positive ($>3\sigma$).
        \item \emph{Not \rev{decaying} in any band}: In each band, the last detection cannot be $>\!\!1\sigma$ below the maximum measured flux, and no point should be $>\!\!1\sigma$ below the immediately preceding one.
        \item \rev{This window must contain at least 5 detections in total, with at least one in each band, and no more than one measurement with negative flux.}
    \end{itemize}

    \item \rev{Historical window:} To exclude objects with persistent long-term activity, we considered only objects without detections in this time window.

    \item \rev{Pre-historical window:} To further restrict the contamination from objects exhibiting persistent large amplitude variability, while still preserving the potential presence of repeated transient flashes, we allowed at most one detection with negative flux in this window. This criterion has proved to be efficient in suppressing objects which historically exhibit large variability, given the current ZTF data, but further improvements on this criterion will be subject of a subsequent version of the module.
    
\end{enumerate}


\subsection{Feature extraction}
\label{sec:features}

Every rising interval, as defined above \rev{in (i)}, was submitted to a lightcurve fit using a simple parametric function within the \textsc{Rainbow} \citep{russeil2024} framework. This approach assumes the transient has a blackbody spectrum and describes its multi-color evolution through two independent parametric models, one for the bolometric flux and another for the temperature. The optimized parameters constitute a highly discriminative feature set, which has been proven effective in separating transient classes \citep{fraga2024}. In particular, \cite{russeil2024} shows that among various transients, TDE classification benefits the most from \textsc{Rainbow} features, since TDE spectra \rev{are} generally well-approximated by a blackbody.

We chose to model the bolometric flux evolution of the rising lightcurves with a simple logistic (sigmoid) function of the form:
\begin{equation}
f(t) = \frac{{\rm \texttt{A}}}{ 1+ \exp( -\frac{t-t_{0}}{{\rm \tau_{rise}}})} 
\label{eq:fit-sigmoid}
\end{equation}

\noindent where \texttt{$\tau_{rise}$} is the characteristic time of rise, \texttt{A} is the amplitude and $t_0$ corresponds to the time of half maximum. We also considered a constant temperature value throughout the rise \rev{$T(t) = T.$}

Such constant temperature assumption is often used when modeling TDEs as it is consistent with optical observations, specially during the rise.
Indeed, while a slight temperature evolution may be observed in later phases, the rise typically exhibits little to no temperature evolution \citep[see e.g.][]{hung2017revisiting, van2020optical}. 
Overall, our \textsc{Rainbow} model has only 4 free parameters. Such minimalist description enables early fits, requiring \rev{a minimum of} only 5 data points, while being sufficient to describe the rising part of most transients lightcurves. 
The lower panel of Figure~\ref{fig:rainbow_example} shows an example of such fit. 

The \textsc{Rainbow} package returns the following features from the fit, alongside their covariances: \texttt{reference\_time} ($t_{0}$), \texttt{rise\_time} ($\tau_{rise}$), \texttt{temperature} ($T$), \texttt{amplitude} ($A$) and \texttt{r\_chisq} (reduced $\chi^2$). We also compute a value reflecting the sigmoid compression, that is used to filter the data (see Section \ref{sec:filtering_labeling}). We define it as \texttt{norm\_rel\_reference\_time} \rev{$= (t_{0} - t_{last}) / \tau_{rise}$}, with $t_{last}$ the time of last detection.

From this set we include only 4 relevant features to train the classifier. We select $T$ and $\tau_{rise}$, which constitute the only physically discriminating features. In addition we use the uncertainty associated to $t_{0}$, \rev{\texttt{e\_reference\_time}}, as an extra informative feature that reflects the degree to which the fit is constrained. Finally, we included the mean angular distance (in pixels) to the nearest object in the ZTF reference image, \texttt{distnr}. This value is used as a proxy for the transient’s offset from the center of its host galaxy, serving as an indicator of how nuclear the event is. When the host galaxy is well resolved, this feature can be particularly helpful in rejecting some off-nuclear transients such as SNe.
However, \texttt{distnr} may become unreliable when the host is not detected in its quiescence state, leading to associations with unrelated nearby objects and, consequently, to meaningless \texttt{distnr} values. To evaluate the usefulness of this feature, we trained the classifier \rev{in two variants -- ``nuclear'' with it, and ``broad'' without including it}.

\subsection{Filtering and labeling}
\label{sec:filtering_labeling}

Prior to the training and classification, we applied an additional filtering step based on the output of the \textsc{Rainbow} fit. Specifically, to avoid very poor fits (which would correspond to data clearly inconsistent with the single-temperature sigmoid model) we excluded the entries with \texttt{r\_chisq} greater than 10.
The entries with signal-to-noise ratios (defined as the parameter value divided by its uncertainty) lower than 1.5 in either rise time or temperature were also excluded.
We also required \texttt{norm\_rel\_reference\_time} to be
between -10 and 0, to exclude overly compressed sigmoid fits (lower bound), as well as highly unconstrained sigmoid shapes whose center lies well after the last alert (upper bound).

For training, we used the subset of all objects that are positionally coincident with the entries in the Transient Name Server\footnote{\url{https://www.wis-tns.org/}} (TNS) as of Jun 1st, 2025, to which we apply the processing presented in Section \ref{subsec:prepro}, \ref{sec:features} and \ref{sec:filtering_labeling}. In constructing a binary classifier, we constructed a set of positives (TDEs) as well as a set of negatives (others).

For the positives, we selected all TNS entries classified as TDEs, irrespective of sub-types, and added a few TDEs reported in \citet{hammerstein2022final} whose classification were not available in TNS at the time of writing. Since there is significant bias in TNS classifications towards supernovae, we decided to label as negatives all TNS objects that are not classified as TDEs, including 4823 that lack any classification. This is justified by the rarity of TDEs among all other possible types of impostors (AGN activity, cataclysmic variables, novae, etc.) which typically remain unclassified in TNS.

\begin{figure}
    \centering
    \includegraphics[width=1\linewidth]{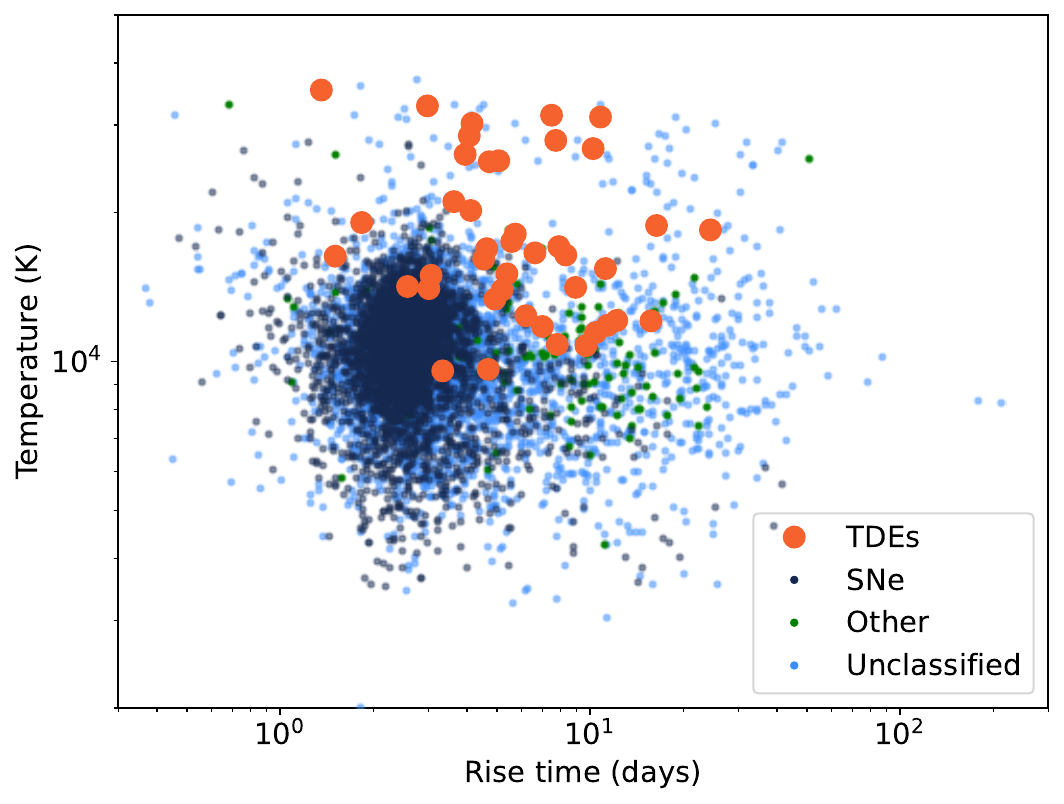}
    \caption{Temperature versus rise time parameter space for the training dataset, showing the imbalance between TDEs and other classes reported in TNS.
    }
    \label{fig:temp_vs_risetime}
\end{figure}

This results in a training dataset with 8865 entries, of which only 42 are positives (known TDEs). 
Figure~\ref{fig:temp_vs_risetime} shows the values of temperature and rise time obtained from \textsc{Rainbow} for all entries in the dataset. 
TDEs tend to exhibit higher temperatures and longer rise time values compared to the bulk of SNe, highlighting the discriminative power of these physically-motivated features\footnote{Note that these rise time values correspond to $\tau_{rise}$ in Equation~\ref{eq:fit-sigmoid}, which, by definition, is shorter than the full duration of the observed rising time, explaining the comparatively small values.}. 
However, the classification task remains challenging: the small number of positives and their relatively broad distribution in the feature space, combined with a significant contamination from the dominant negative class, hinder the classification task.

\subsection{Classifier}
\label{sec:classifier}

To tackle the issue of our extremely imbalanced dataset, we aimed at building a classifier capable of efficiently generalizing the feature behavior of a few positives, thus producing predictions over a feature space mostly occupied by negatives. For this, we used the extreme gradient boosting algorithm implemented in the \textsc{XGBoost}\footnote{XGBoost: eXtreme Gradient Boosting library, \url{https://xgboost.readthedocs.io/en/release_3.0.0/}.} package \citep{xgboost}.
This method combines the predictions of several weak learners, in the form of decision trees, to create a stronger learner.

As the dataset is highly imbalanced, we oversampled the positives using Synthetic Minority Over-sampling Technique\footnote{We used SMOTE implementation from \textsc{imbalanced-learn} package, \url{https://github.com/scikit-learn-contrib/imbalanced-learn}.} (SMOTE, \citealt{smote}). In order to better cover the region with synthetic positives we applied the oversampling in two steps, first interpolating the known TDEs up to half of the negatives, and then interpolating the results to fully balance the dataset. 
To minimize over fitting, we fixed the hyperparameter defining maximal depth of individual decision trees (\texttt{max\_depth}~$=3$), forcing the trees to be shallow. It ensures the splitting of the parameter space into broad regions, thus significantly suppressing overfitting.
For the optimization of the remaining hyperparameters we used stratified 5-fold cross-validation, maximizing the \texttt{F2} score defined as:
\begin{equation}
    \mbox{F2} =  \frac{5 \cdot \mbox{precision}\cdot\mbox{recall}}{4\cdot\mbox{precision} + \mbox{recall}}
\end{equation}

This score, in contrast to the standard \texttt{F1} score, places greater emphasis on recall than on precision, which improves the extrapolation capacities of the classifier despite the heavily imbalanced classes. Thus, favoring the reduction of false negatives (TDEs classified as others) at the cost of obtaining more false positives (others classified as TDEs).

\begin{table}[h!]
\caption{Hyperparameters for the XGBoost classifier, and corresponding final scores, for ``nuclear'' and ``broad'' models.
}
\label{tab:hyperparameters}
\centering          
\footnotesize

\begin{tabular}{lcc}
\hline\hline
Name & \multicolumn{2}{c}{Value}\\
& \ \ \ \ \ \ Nuclear\ \ \ \ \ \ \ & \ \ \ \ \ \ Broad\ \ \ \ \ \  \\
\hline
\\
\multicolumn{3}{c}{Manually set}\\
\hline

\texttt{max\_depth} & 3 & 3 \\

\\
\multicolumn{3}{c}{Optimized for \texttt{F2} score}\\
\hline

\texttt{n\_estimators} & 200 & 200 \\ 
\texttt{subsample} & 0.8 & 0.8 \\
\texttt{reg\_lambda} & 1 & 1 \\
\texttt{reg\_alpha} & 1 & 1 \\
\texttt{learning\_rate} & 0.06 & 0.06 \\
\texttt{colsample\_bytree} & 1.0 & 1.0 \\
\texttt{min\_child\_weight} & 5 & 5 \\

\\
\multicolumn{3}{c}{Final scores}\\
\hline

\texttt{F2} score & 0.37 & 0.25 \\
\texttt{F1} score & 0.20 & 0.13 \\
Precision & 0.12 & 0.07 \\
Recall & 0.76 & 0.74 \\

\hline
\end{tabular}
\tablefoot{The hyperparameters were optimized by a stratified 5-fold cross-validation routine, while the final scores are derived by the Leave-One-Out method.}
\end{table}

\begin{figure}[h!]
    \centering
    \includegraphics[width=0.48\columnwidth]{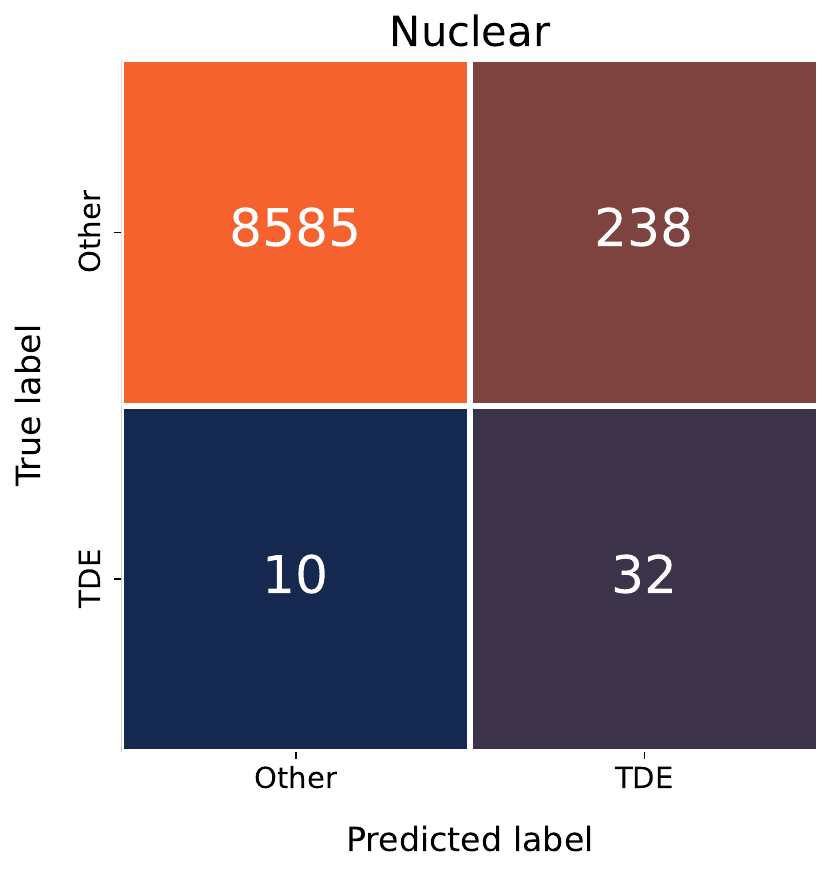}
    \includegraphics[width=0.492\columnwidth]{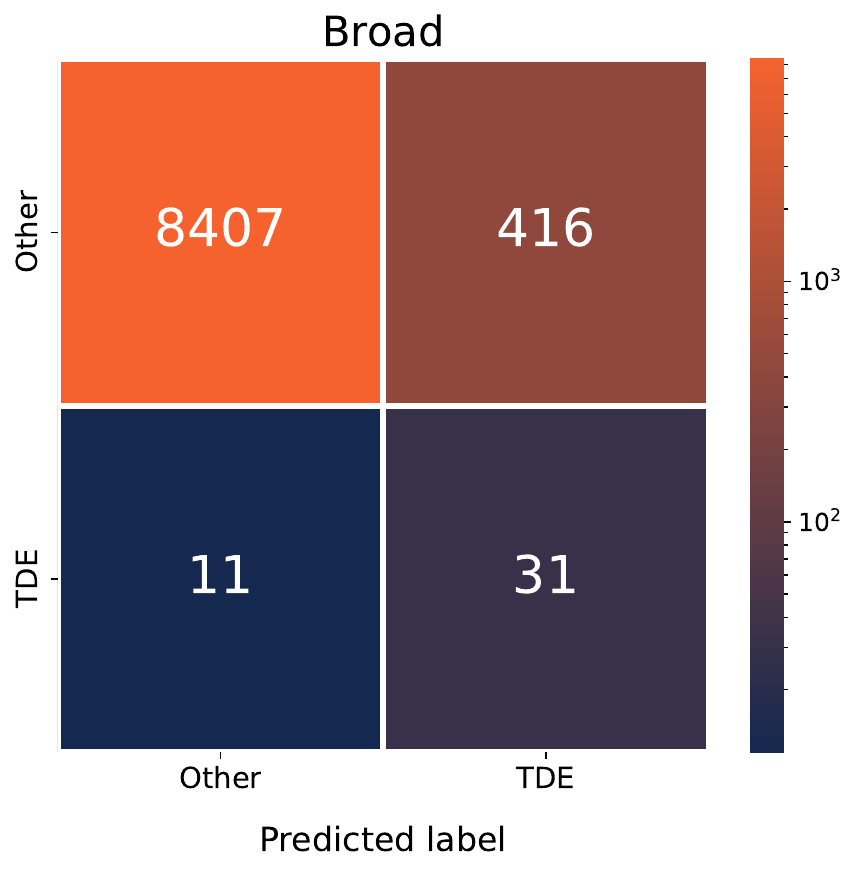}
    \includegraphics[width=0.49\columnwidth]{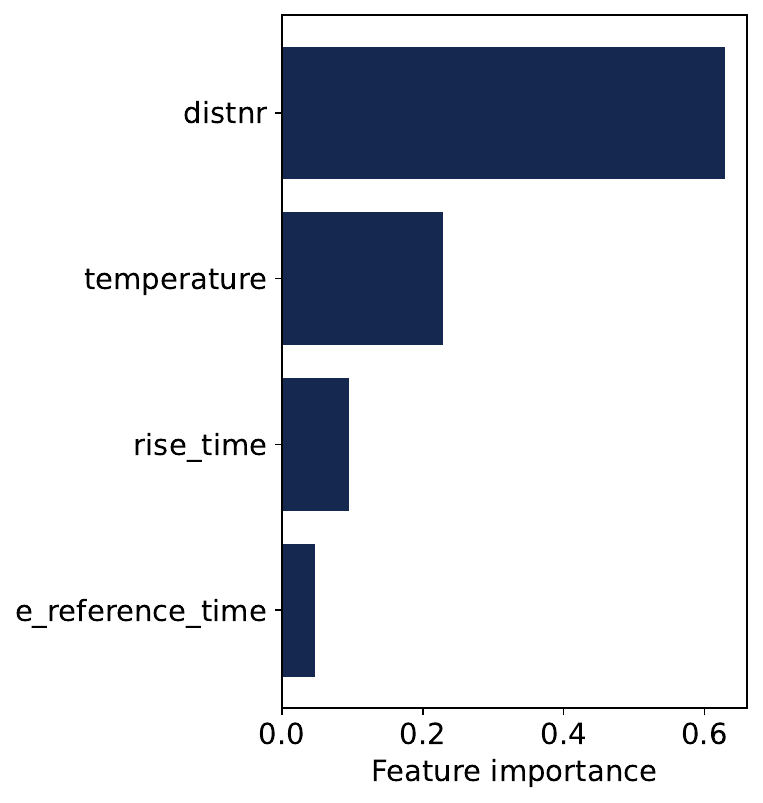}
    \includegraphics[width=0.49\columnwidth]{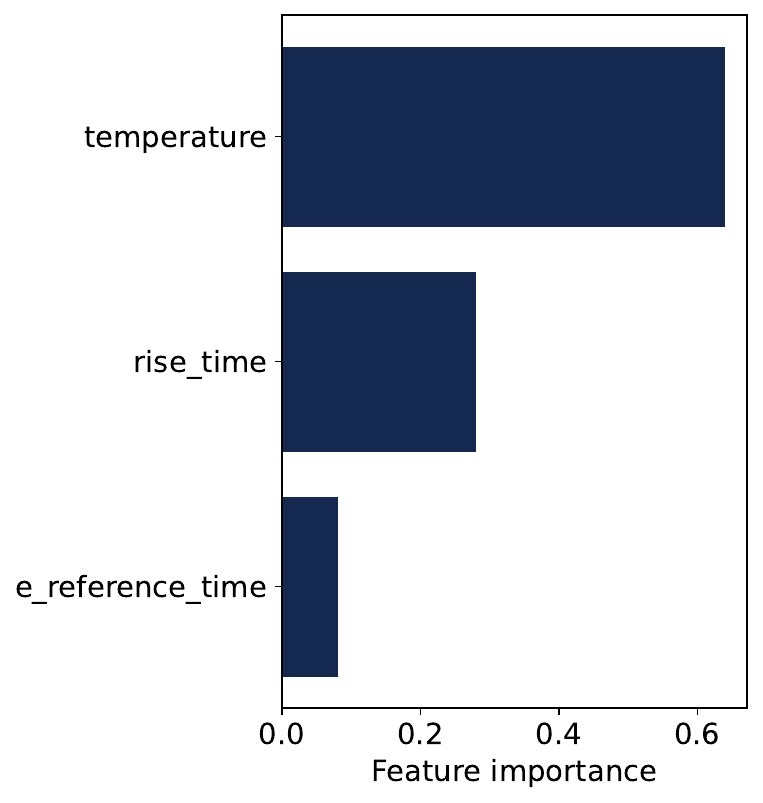}
    \caption{Confusion matrix built using Leave-One-Out cross-validation (upper panels) and relative feature importance (lower panels) for the \rev{``nuclear''} (left panels) and \rev{``broad''} (right panels) models.}
    \label{fig:confusion_matrix}
\end{figure}

Together with hyperparameter optimization \rev{(Table \ref{tab:hyperparameters})}, we performed iterative pruning of features which are not important for classification results, as these may impact performances on independent data sets, and are likely to be highly correlated with other features\footnote{E.g. the bolometric definition of amplitude in \textsc{Rainbow} (see Eq.~\ref{eq:fit-sigmoid}) makes it highly correlated with the temperature given the measurements in a fixed spectral band, thus reducing its effectiveness for the classification.}.
Thus, as detailed in Section \ref{sec:features}, the features fed to the machine learning model are \texttt{temperature}, \texttt{rise\_time}, \texttt{e\_reference\_time} and optionally \texttt{distnr} \rev{(for the ``nuclear'' classifier)}. 
Table~\ref{tab:hyperparameters} shows the final hyperparameter values used in the models.

\begin{figure*}[h!]
    \centering
    \includegraphics[width=2\columnwidth]{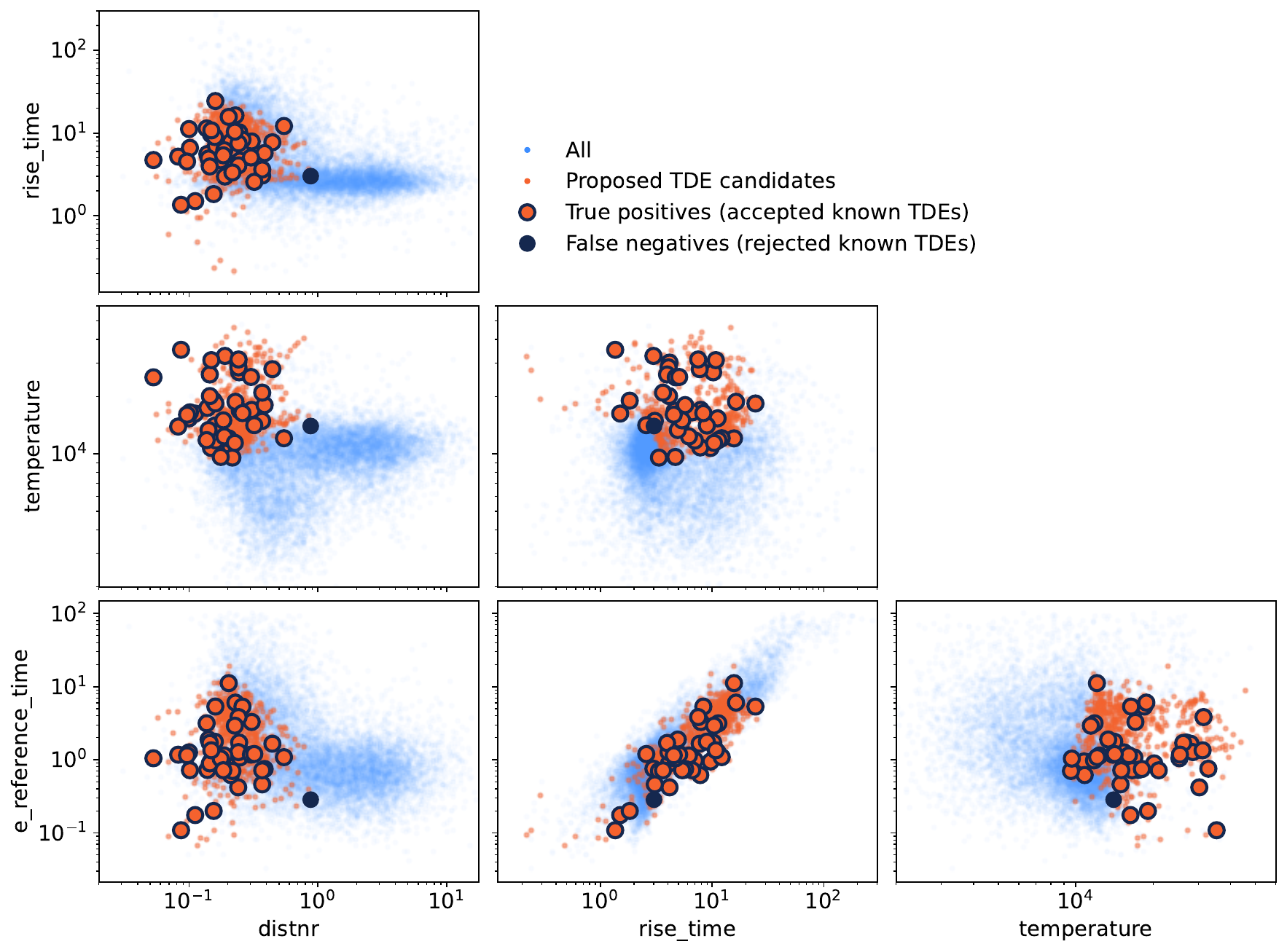}
    \caption{Corner plot showing the complete feature space for the ``nuclear'' model \rev{on the whole training dataset, and applied to all lightcurves}. Small blue dots show the positions of all points in the dataset, orange ones mark the TDE candidates proposed by the classifier. Orange and dark blue circles show the positions of known TDEs that are accepted and rejected by the classifier, respectively. \rev{Note that, contrary to the results shown by the Leave-One-Out cross-validation in Figure~\ref{fig:confusion_matrix}, only one known TDE is not identified by the model in this test because all these TDEs are included in the training set. This outlier is the off-nuclear optical TDE AT2024tvd.}
    }
    \label{fig:cornerplot}
\end{figure*}

\section{Results}
\label{sec:results}

We present results from validation of the pipeline using labels obtained from the literature (Section \ref{subsec:Val}), as well as a selection of interesting candidates found during the development of this module (Section \ref{subsec:candidates}).

\subsection{Validation}
\label{subsec:Val}

To assess the classifier's performance with optimized parameters when trained on the full dataset, we performed a run of Leave-One-Out cross-validation by excluding, one by one, a single point from the original training set, oversampling and training on the remaining data.
The final scores are listed in Table~\ref{tab:hyperparameters}, while the final confusion matrix and relative feature importance are shown in Figure~\ref{fig:confusion_matrix}. Taking into account the imbalance between the two classes, the confusion matrices confirm the success of our initial goal of minimizing the number of false negatives. In both scenarios, the majority of TDEs were identified by the algorithm, 32 (76\%) 
when \texttt{distnr} is available (``nuclear'' classifier)
and 31 (74\%) otherwise. 
\new{Despite its minor impact on recall, this additional feature significantly improves precision by enabling the rejection of many contaminants, namely off-nuclear transients such as SNe.}
The lower panels of the same figure showcase the dominant importance of this additional feature.
The order of importance for the remaining features is maintained when this distance is not available. 

\new{Although the \texttt{distnr} feature aids in filtering out contaminants, it is not essential for a successful classification, as reflected in its minimal impact on recall, and may even hinder the identification of rare off-nuclear TDEs.} Notably, recent discoveries such as the off-nuclear TDE AT2024tvd \citep{yao2025offnuclear_AT2024tvd} and the TDE-like event AT2024puz \citep{offnuclear_AT2024puz} suggest that relying too heavily on this criterion could be overly restrictive. To balance these considerations, we opted to use both classifiers in parallel, combining their outputs and further investigating promising candidates identified by either.

To evaluate the classifiers on archival data, we applied them to the full dataset described in Section~\ref{sec:module_design} that covers observations between Nov 2019 and Feb 2025, including objects not listed in TNS.
Figure~\ref{fig:cornerplot} presents the feature space obtained with the final ``nuclear'' model 
for this full dataset. As shown, the proposed candidates (orange dots) occupy the same region of feature space as the known TDEs (circles), while rejected objects (blue dots) lie elsewhere, showing the usefulness of the chosen features. Interestingly, the only\footnote{Most known TDEs were identified as they were all used to train the model.} known TDE not flagged as such in this test run (dark blue circle) is the aforementioned off-nuclear TDE AT2024tvd. This is a direct consequence of its \texttt{distnr} value, which is noticeably bigger than the other TDEs.
This further justifies also executing the ``broad'' model without the \texttt{distnr} feature so results enclose a large variety of candidates. 
\rev{This also highlights that the model is not simply memorizing the training set, but instead generalizing beyond it.}

Upon inspection of the reported TDE candidates, we identified several promising TDEs (see Section~\ref{subsec:candidates}). This test also clearly indicated that the primary contaminants are AGNs, whose typical variability might result in rising lightcurve segments that resemble the early phases of TDEs.

\subsection{Some identified candidates}
\label{subsec:candidates}

\begin{figure*}
    \centering
    \includegraphics[width=1\columnwidth]{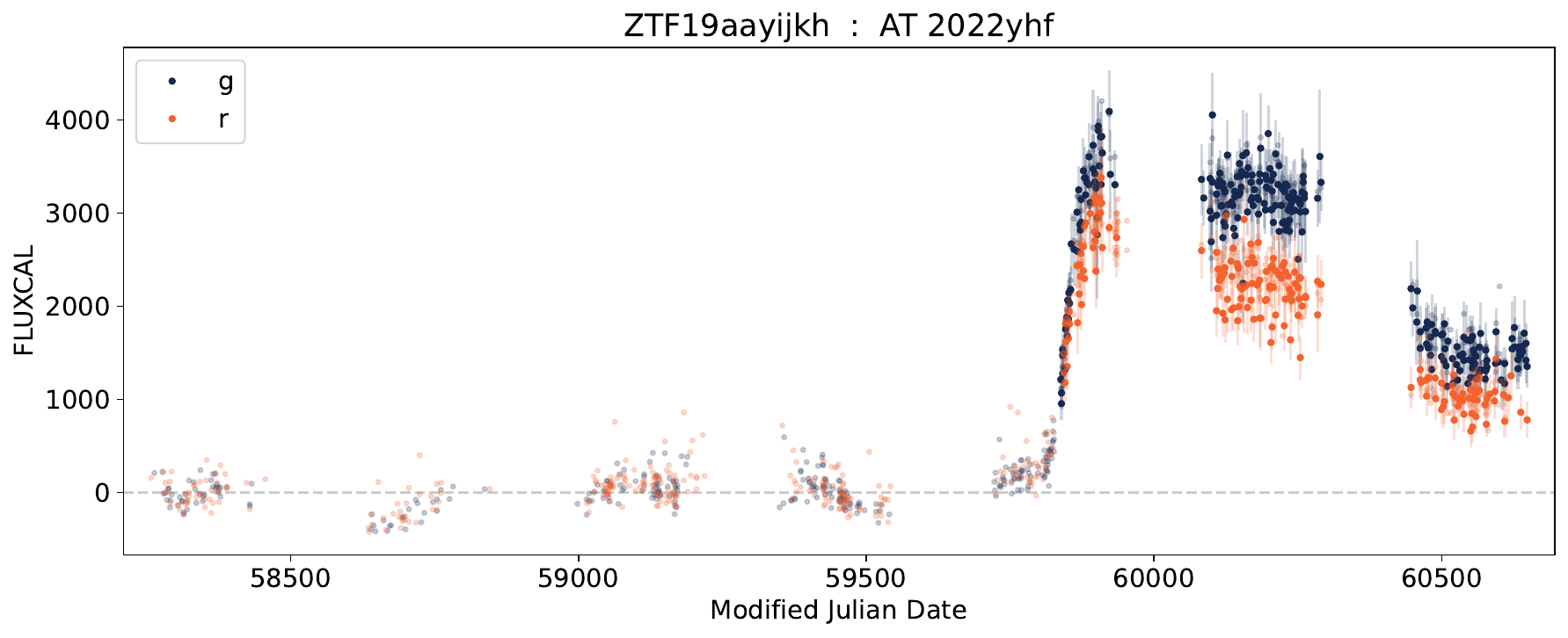}
    \includegraphics[width=1\columnwidth]{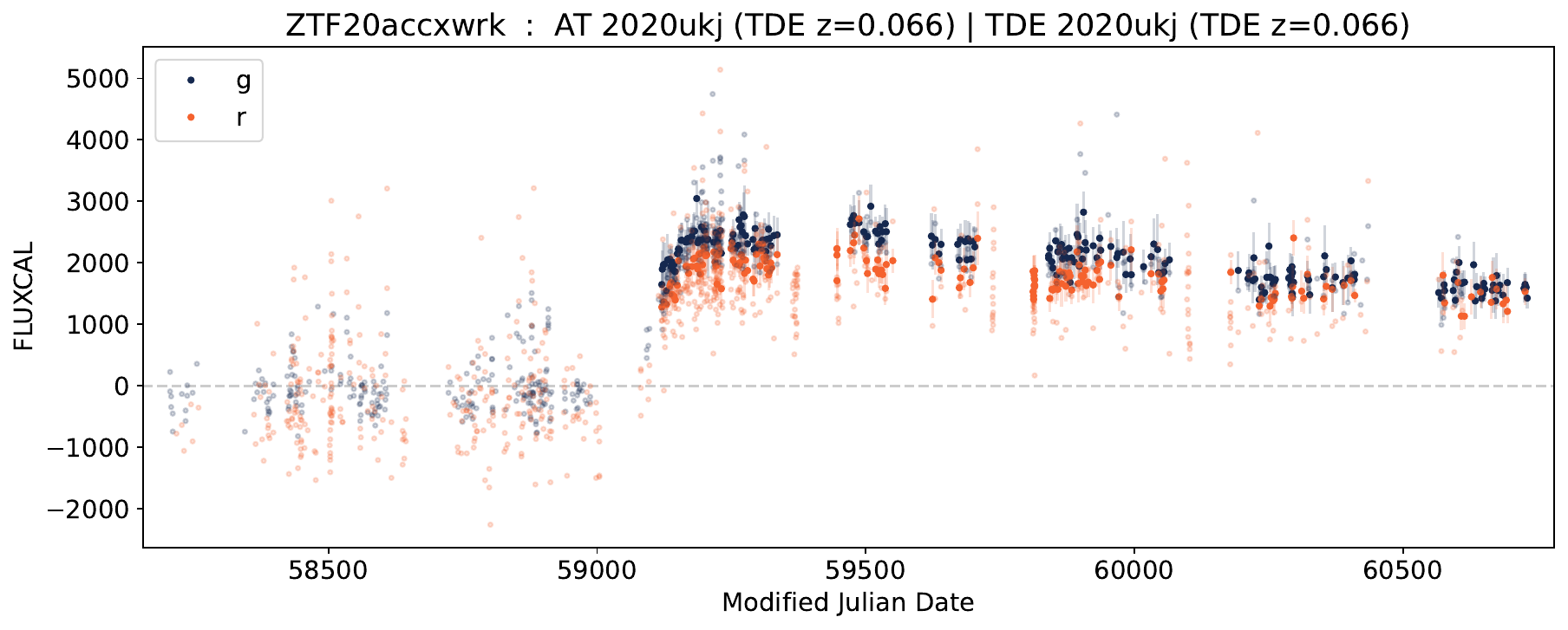}
    \includegraphics[width=1\columnwidth]{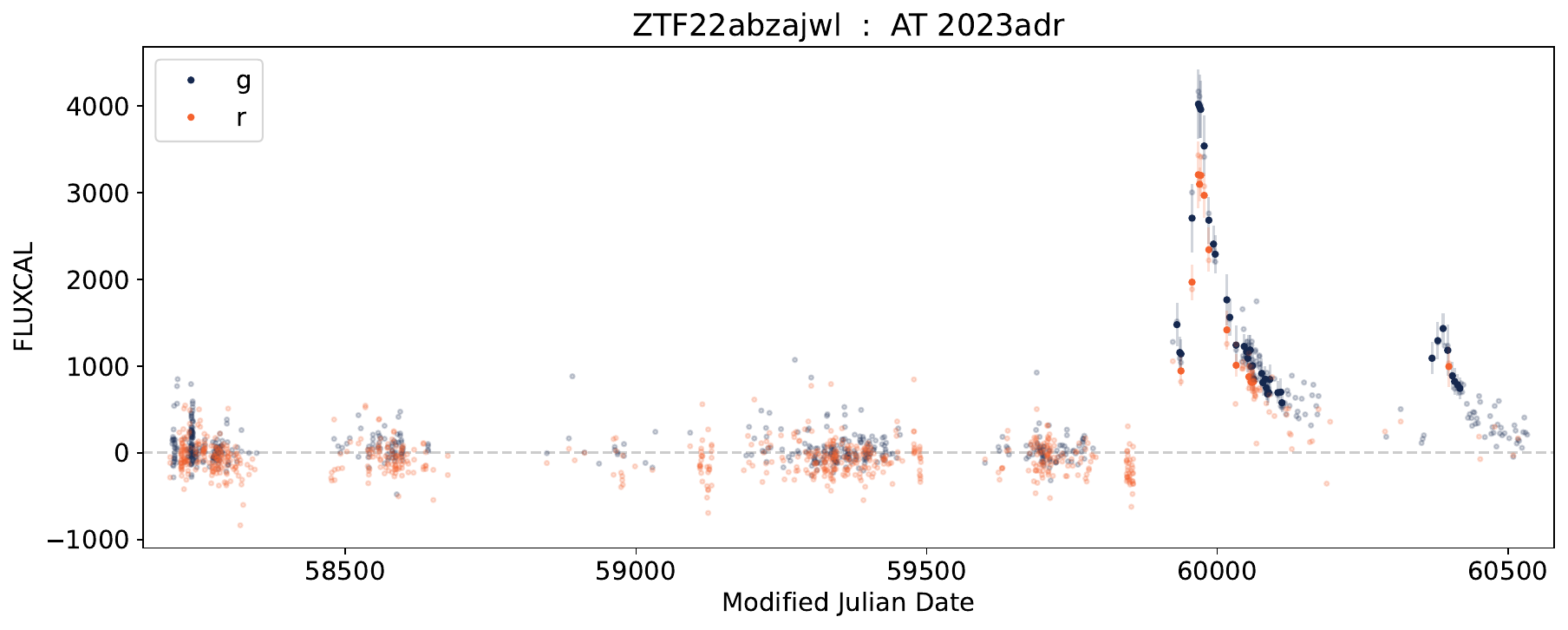}
    \includegraphics[width=1\columnwidth]{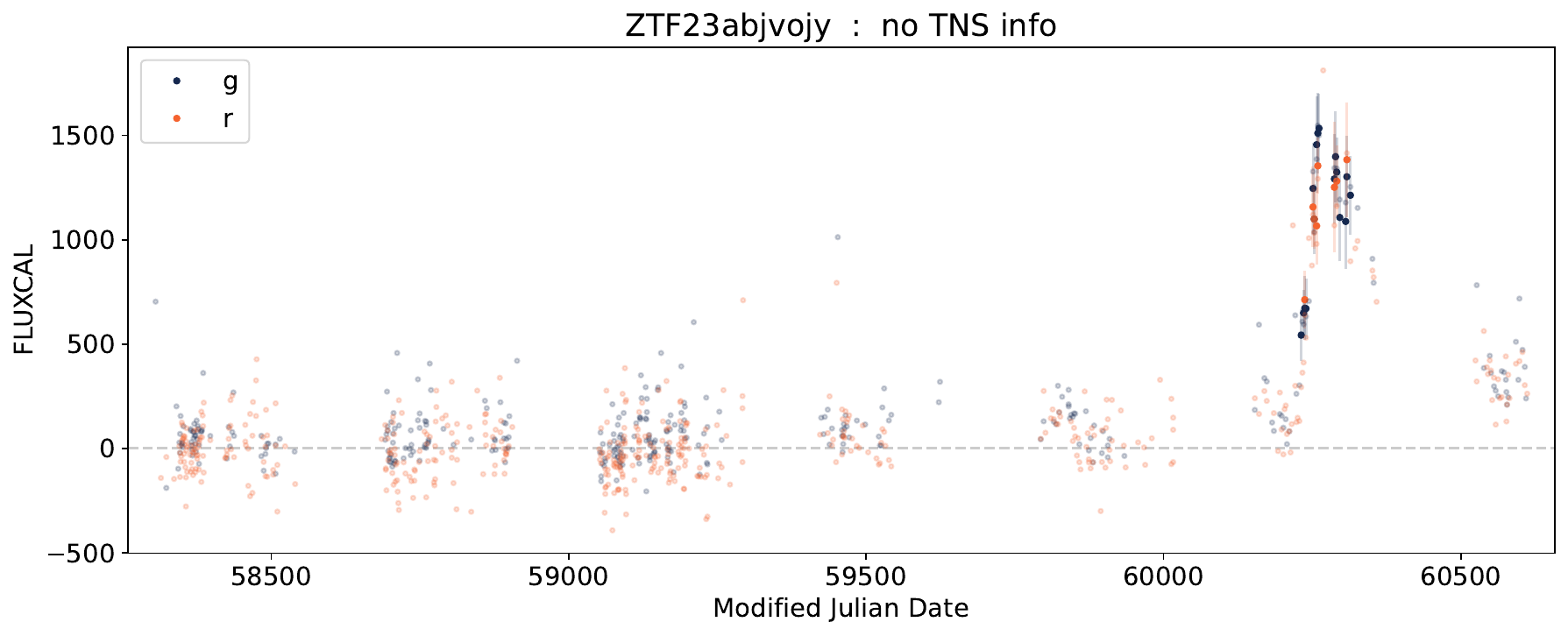}
    \caption{lightcurves of four TDE candidates identified by the module. Full dots represent detections extracted from \fink\ alerts, while semi-transparent ones are extracted from data-release photometry.
    }
    \label{fig:candidates}
\end{figure*}

Throughout the different stages of the module development, several interesting transients were identified as potential TDE candidates by analyzing archival data. A detailed analysis of these transients will be provided in the companion paper, Quintin \textit{et al.} - in prep. Here, as an illustration, we present a summary of four of these objects, with their corresponding long-term ZTF lightcurves displayed in Figure~\ref{fig:candidates}. Full dots represent alerts, while translucid dots are extracted from ZTF data-release photometry\rev{\footnote{\rev{Data-release photometry refers to ZTF detections available in public data releases but not necessarily triggered as real-time alerts.}}}, retrieved using the SNAD API \citep{Malanchev2023}. \\

\subsubsection{ZTF19aayijkh / AT2022yhf}

This transient\footnote{\url{https://fink-portal.org/ZTF19aayijkh}} is hosted by a blue active galaxy located at a photometric redshift of $z\sim0.56\pm0.03$ \citep{almeida2023eighteenth}, classified as a Quasar (QSO) by Gaia DR3 \citep{gaia_collaboration_gaia_2023}. The host exhibits QSO-like variability on timescales of $\sim6$ months.

On top of this variability, a significant brightening event starting on September 2022 was identified by our module as a potential TDE. The flare is characterized by a slow $\sim3-4$ months rise followed by a steady decay that continues for more than two years after peak brightness (upper left panel of Figure~\ref{fig:candidates}). 
The event is significantly bluer than the quiescent state, reaching its bluest at peak with little to no reddening observed during the subsequent decay.
The transient's apparent magnitude at the peak and the host redshift correspond to luminosity values of the order of $\sim10^{44}\,erg/s$ in both bands.
This luminosity, combined with the observed color evolution, is consistent with a TDE occurring within the QSO.
In addition, data from the Wide-field Infrared Survey Explorer (WISE or NeoWISE, \citealt{wright2010wide}) reveal hints of an associated infrared echo emerging in May 2025, which corroborates the TDE hypothesis.

\subsubsection{ZTF20accxwrk /AT2020ukj}

This object\footnote{\url{https://fink-portal.org/ZTF20accxwrk}} is a very long-lived transient that has been steadily decaying since its peak on December 2020, as depicted in the upper right panel of Figure~\ref{fig:candidates}. Its host, with photometric redshift of $z\sim0.089$ \citep{duncan_all-purpose_2022}, was not active beforehand and exhibited a red color ($g-r\sim1$). 

The 2020 brightening features a bluer color (reaching $g-r\sim0.6$), which has been reddening steadily since, without yet reaching the color of its quiescent state.
This source exhibits striking similarities to ZTF19acnskyy \citep{sanchez2024sdss1335} in terms of timescales, particularly in its exceptionally slow decay compared to the slowest known TDEs. This supports a similar physical interpretation for both objects: either an unusually long-duration TDE, or an AGN in the process of activation.

We acquired a spectrum using the 2.5m telescope from the Caucasian Mountain Observatory (CMO, \citealt{shatsky2020caucasian}), which supports a TDE classification at $z=0.066$ \citep{tde_2020ukj}. However, the observed luminosity leads to an energy budget possibly too large for the disruption of a single star, favoring the interpretation of a starting AGN. We are currently monitoring this source in X-rays, to investigate its multi-wavelength behavior.

\subsubsection{ZTF22abzajwl / AT2023adr}

This object\footnote{\url{https://fink-portal.org/ZTF22abzajwl}} presented a flare starting in December 2022 which had been first identified as a potential Superluminous SN by \citet{perley2023ztf}, and then as a potential TDE by \citet{aleo2024anomaly}.
A second fainter flare occurred in February 2024, as seen in the lower left panel of Figure~\ref{fig:candidates}. The ePESSTO+ group obtained a spectrum for this object while searching for SNe, locating it at $z\sim0.131$ using the clear narrow-lines \citep{shlentsova2024epessto}. The spectrum shows no strong metal lines or very broad emission lines which could be expected from SNe, and the smoothly evolving lightcurve tends to exclude circumstellar medium (CSM) interaction.

Upon identification of this candidate by our module, the presence of both TDE-like flares and the non-SN-like spectrum led us to classify it as a TDE, and in particular a partially repeated TDE \citep{llamaslanza2024_pTDE}. This discovery adds a new source to the relatively short list of known or candidate repeated TDEs, which include ASASSN-14ko \citep{payne2021asassn, payne2023chandra}, eRASSt~J045650.3–203750 \citep{liu2023deciphering, liu2024rapid}, AT2018fyk \citep{wevers2023live}, RX~J133157.6–324319.7 \citep{malyali2023rebrightening}, AT~2020vdq \citep{somalwar2025first} and AT~2022dbl \citep{lin2024unluckiest}.

\subsubsection{ZTF23abjvojy}

A brightening event\footnote{\url{https://fink-portal.org/ZTF23abjvojy}} starting in October 2023 was flagged by our module as a TDE candidate. The host is a spectroscopically confirmed broadline QSO at $z\sim0.277$ \citep{almeida2023eighteenth}, with hints of quiescent level variability. The identified flare, however, represents a significant excess from this baseline, as shown in the lower right panel of Figure~\ref{fig:candidates},
and does not appear to be the typical QSO red-noise variability.

The source rose with a timescale of $\sim25$ days becoming significantly bluer, and decayed in less than 200 days, with no significant color change in the decay.
Additionally, a clear infrared echo of this transient is found in the NeoWISE lightcurve.
Interestingly, another even brighter flare from this galaxy is visible more than 15 years earlier in data from the Catalina real-time transient survey (CRTS, \citealt{drake2009first}) starting to rise in January 2008 and slowly decaying over a decade. 
The timing and luminosity properties of both brightening episodes are consistent with a TDE or an AGN flare. In particular, the significant asymmetry of the first peak is not consistent with standard AGN variability. Although it is not possible to conclude on the nature of these bursts without spectral observations, this source is a good candidate to join a small group of repeated TDE-like flares in AGNs, along with AT2019aalc \citep{milan_veres_back_2024}, AT2021aeuk \citep{sun_at2021aeuk_2025}, and IRAS F01004-2237 \citep{sun_recurring_2024}.

\rev{
\subsection{Early identification performance}}
\label{sec:early_detection}

\begin{figure}[h]
    \centering
    \includegraphics[width=1\linewidth]{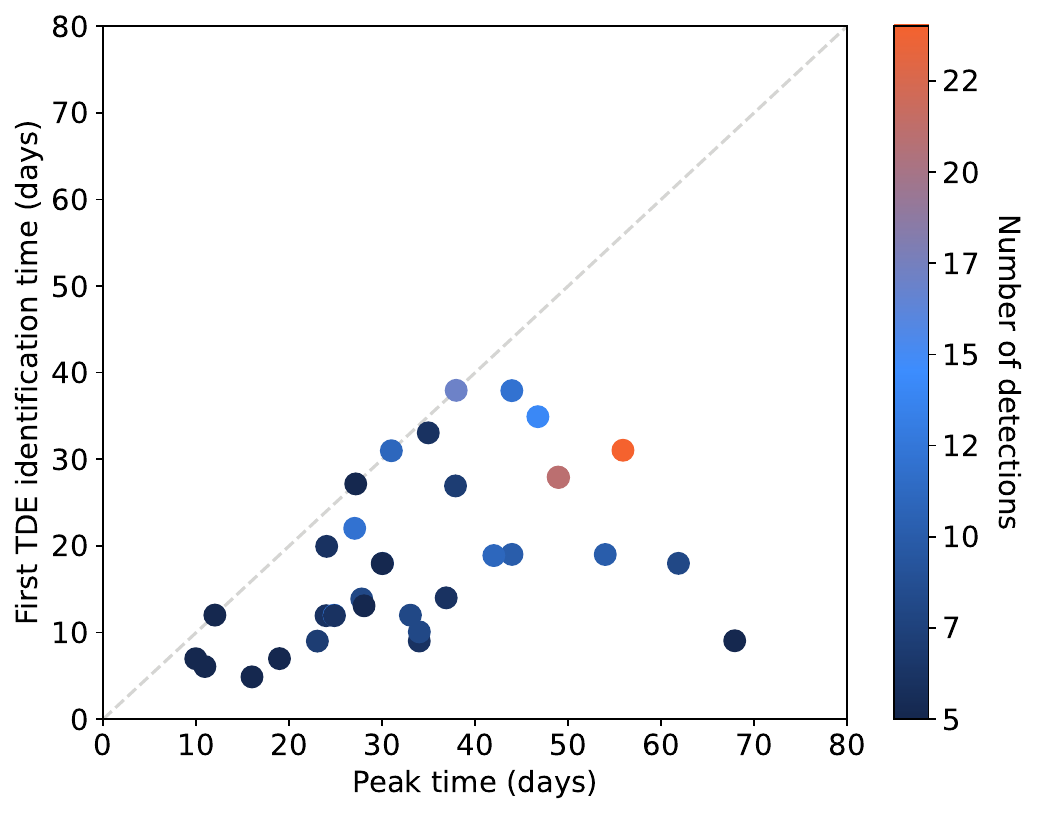}
    \caption{Time of the first alert that led to a TDE candidate identification as a function of the peak-flux time for each known TDE correctly classified by our module. 
    Both times are measured with respect to first detection within 100 days prior to the peak. The color scale illustrates the number of detections within this window culminating in a true positive classification.}
    \label{fig:first_score}
\end{figure}

Considering that the main objective of this work is to identify candidates as early as possible to enable their follow-up, it is relevant to assess how early in their rise we may actually identify them.
To do so, we evaluated each known TDE in our sample using the Leave-One-Out approach (see Section~\ref{subsec:Val}) to train the ``nuclear'' classifier on all the dataset except one TDE event at a time, and then applying it for all segments of the excluded event's lightcurve. Such procedure ensures the resulting models in each iteration are as similar as possible to the final model, presented in Section~\ref{sec:classifier}, while avoiding training on the test object.
From this evaluation, we extract for each candidate the time of the first alert classified as a rising TDE.

Figure~\ref{fig:first_score} shows the time of first identification as a function of the time of maximum flux for each TDE candidate. Both quantities are measured with respect to the first detection within the 100-day window preceding the peak, which itself is defined as the last data point satisfying \textit{the rising and not \rev{decaying}} criteria described in Section~\ref{subsec:prepro}. Out of the 42 TDEs in the sample, 34 (81\%) were correctly classified at least once during their rise, including 4 that were only flagged at the peak (data points on the diagonal). 
The latter might demand a more rapid follow-up response, depending on their brightness and how fast they decay. Nevertheless, the remaining cases were identified earlier, with the delays predominantly dominated not by the classifier itself but instead by preliminary selection criteria like flux availability in both bands.
Half of these TDEs were identified before reaching 52\% of their rising phase, and a quarter of them were identified as early as a 38\% of their rise.
Such early identification enables potential monitoring of the rest of the rising phase and provides enough time to potentially perform follow-up observations at their peak.

\section{Discussion}
\label{sec:discussion}

\subsection{Classifier design choices}

Probably the biggest challenge in this work was conceiving a classifier able to learn, and generalize specific characteristics from the small number of known TDEs available for training. Thus, enabling the identification of less-canonical or previously unobserved TDE-like events, making the module sensitive to a broader spectrum of behaviors consistent with theoretical expectations. 

Moreover, we prioritized completeness (recall) over purity (precision). While this increases the likelihood of detecting true TDEs, it also leads to a higher rate of false positives, requiring a subsequent manual vetting step. 

Aligned with this philosophy, we decided to loosen certain criteria that could be applied to exclude AGNs, such as applying stricter cuts on historical activity or removing AGNs based on catalog cross-matches. This decision was motivated by the possibility of capturing repeated TDEs as well as TDEs occurring within AGNs, which are theoretically expected to be common but observationally challenging to identify on top of the usual AGN variability.
Naturally, this approach increases the number of false positives as AGNs are one of the main contaminants, but it broadens the range of TDE characteristics accessible to this module.
In future versions, we plan to refine this approach by incorporating a comparison between the variability observed in the rising phase and the historical baseline activity, allowing us to better distinguish TDE flares from typical AGN variability and thereby reduce contamination while still preserving completeness.
Another example of this completeness-driven philosophy is retaining the ``Broad'' model, which does not include the \texttt{distnr} feature despite its usefulness in reducing contaminants, in order to remain sensitive to rare off-nuclear TDEs.

\subsection{\fink\ filter implementation}
\label{sec:filter}

Once the classifier was finalized, we proceeded to implement it into the \fink\ broker to enable automatic processing. In this context, beyond classification results, we also aimed for a user-friendly delivery method which would allow astronomers to quickly go through potential candidates and thus make informed follow-up decisions. 
To do so, we implemented a filter within \fink\ that is being applied to the data acquired by the broker after every night of its operation. The filter then selects the candidates and sends them to the dedicated \fink\ community Slack and Telegram\footnotemark{} channels for review.

Since the information contained in individual ZTF alerts is limited to 30 days of photometry and access to historical data beyond that is relatively costly, we had to adjust the methodology outlined in Section~\ref{subsec:prepro}.
We require at least 5 detections, with at least one point in each band, $g$ and $r$, and no more than 1 negative flux in total to these 30 days only. Moreover, we also required these data points to satisfy the \textit{rising and not decaying} criteria as described above. Table~\ref{tab:streaming_stats} provides an overview of how many alerts remain after every specific step of this filtering.

\begin{table}
\caption{Number of alerts on various steps of the processing of ZTF data stream in 84 nights of Jan -- Apr 2025.}
\label{tab:streaming_stats}
\centering          
\footnotesize

\begin{tabular}{lr}
\hline\hline
Step & Number of alerts \\
\hline
\\
\multicolumn{2}{c}{Individual alerts level}
\\
\hline

Received by \fink & 14,833,282 \\
Passed internal quality cuts & 9,813,510 \\ 
After MPC filter & 8,572,610 \\
After SIMBAD type filter & 3,981,797 \\
At least 5 good points & 1,545,771 \\
No more than 1 negative point & 694,212 \\
Both bands have data & 614,572 \\
Galactic $|b|>20$ & 322,362 \\
Rising and not \rev{decaying} & 9,538 \\

\\
\multicolumn{2}{c}{Full lightcurves level}
\\
\hline

Detections prior to 100 days & 3,735 \\
Good quality \textsc{Rainbow} fit & 3,632 \\
Quality cuts on features & 1,452 \\
$\geq10\%$ of probable fits \rev{identified ($p>0.5$)} & 248 \\
\ \ \ \ unique objects & 111 \\
Best fit \rev{identified ($p>0.5$)} & 130 \\
\ \ \ \ unique objects & 45 \\

\hline
\end{tabular}
\end{table}

These steps significantly reduce the number of candidates, down to approximately a hundred per observing night. This number is small enough so that full-scale processing on their historical lightcurves could be performed.
From this point, we applied exactly the same cuts on the number of detections in the \rev{fitting, historical and pre-historical windows (see Figure~\ref{fig:rainbow_example}) as described in Section~\ref{subsec:prepro}.}
We then extracted features using the detections and upper limits inside the fitting window as described in Section~\ref{sec:features}. Finally, we applied quality cuts to these features, as described in Section~\ref{sec:filtering_labeling}.

In order to maximize the detection probability, we adopt an approach that, to some degree, accounts for the statistical uncertainties of the features derived from the lightcurve during the \textsc{Rainbow} fitting process. The fitter reports the best fit values of the parameters together with their covariances. 
\rev{Using this information, we generate multiple feature sets by sampling the distribution of probable parameter values around the best fit consistent with the fit uncertainties.
These sampled feature sets are then passed through the classifier and a candidate is classified as a potential rising TDE if either the best fit set or at least 10\% of the sampled sets are positively classified as such.}

\begin{figure}
    \centering
    \includegraphics[width=1\linewidth]{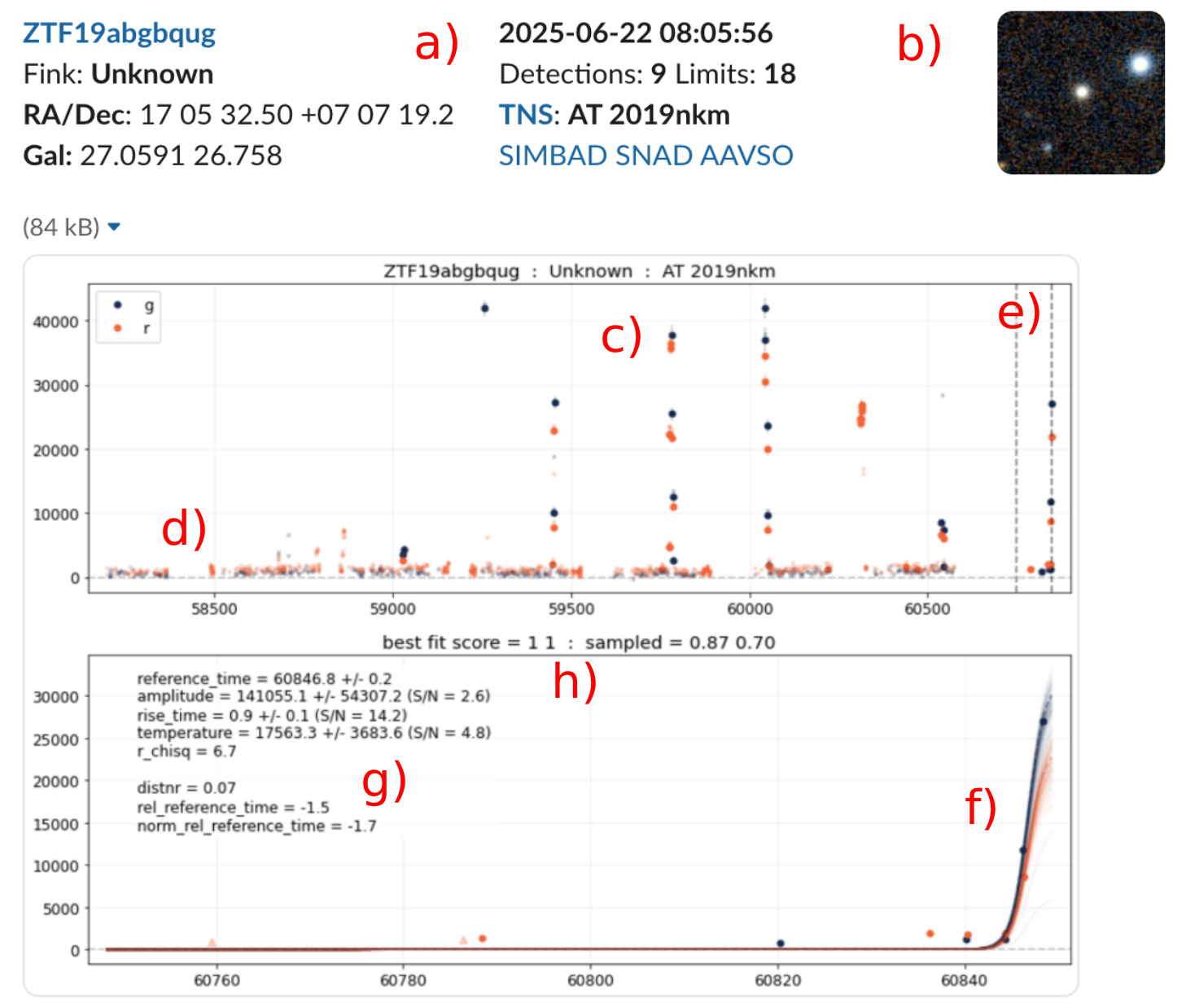}
    \caption{Example information card as produced by the \fink\ filter and posted to dedicated Slack channel \rev{as a candidate. Despite not corresponding to a TDE, we chose to show this object as it allows to clearly illustrate each of the following elements: }  a) information block including the ZTF object identifier, alert timestamp and sky coordinates, along with TNS name and classification when available, and direct links to \fink\ object page and several external services that might contain additional useful data about this object; b) Pan-STARRS cutout image centered at the object position; c) historical lightcurve based on data from alerts and d) latest ZTF data-release photometry; e) the latest 100-day interval; f) best fit curve of the data over last 100 days, plus the probable fits sampled from the covariances of the parameters; g) values and error bars for the fit parameters and additional features; h) binary scores from two classifiers for the best fit, and the fraction of probable fits identified as rising TDE candidates.
    }
    \label{fig:card}
\end{figure}

The candidates are then published to dedicated Slack and Telegram\footnotetext{\url{https://t.me/fink_early_tdes}}\footnotemark[\value{footnote}] channels. 
For every candidate, the filter constructs a compact card-like representation containing a variety of available information, including the historical lightcurves reconstructed from both alerts and latest ZTF data-release photometry \citep{Malanchev2023}, 
along with the supporting data from TNS, and direct links to the \fink\ science portal and several other online resources which might be useful for rapid screening of the transient. An example of such a card is shown in Figure~\ref{fig:card}\rev{, for a lightcurve that was selected by the module as a candidate. Despite not corresponding to a TDE, we chose this object as it allows to clearly illustrate each component of the card, and particularly because it exhibits past flares in the pre-historical window (see Section~\ref{subsec:prepro}), which are accepted by our module.}

\subsection{Other classifiers}

Comparisons among machine learning classifiers performances should be interpreted with caution, as differences in datasets, selection criteria, and metadata often prevent direct equivalence between analyses. Consequently, classifier results are most meaningful when considered within their full methodological context. Nevertheless, we attempt a comparison with the FLEET \citep{Gomez2023} TDE classifiers to provide a general perspective. 

It currently constitutes, to the best of our knowledge, the only other model that attempts to detect early TDE lightcurves. However, we underline that our definitions of early lightcurves are distinct. While we consider any lightcurve still rising to be early, \cite{Gomez2023} defines it as the first 20 days of data after discovery. Hence, a decaying transient that recently appeared could be considered early. From this, FLEET has been trained using ZTF data, and limiting the sample to sources that lie within the footprint of the Pan-STARRS1 (for host galaxy association). From a dataset of 45 TDEs among 4779 transients, it achieves 50 \% purity and 30 \% completeness for P(TDE)> 0.7. To the contrary, our implementation observes a higher completeness for the price of less purity. Indeed, using the \textit{distnr} metadata, it reaches 12 \% purity and 76\% completeness. Therefore we find significantly more TDEs, in particular events challenging to identify with high confidence at early stages. However, this increase is expected to require greater expert involvement to determine which events should be prioritized for follow-up observations. Although it has not been optimized for this purpose, the default 0.5 classification threshold can also be manually raised to increase the purity. For example, by only selecting events classified as TDE with more that 99\% confidence, the classifier reaches a compromise of $\sim 40\%$ for both, purity and completeness.

\subsection{Preparing for the Rubin alert stream}
\label{sec:rubin}

The module we implemented is optimized for the cadence and other properties of the ZTF sky survey. However, the future plan is to adapt it for the Rubin/LSST data stream, which is expected to start producing data in the second half of 2025. 
While its cadence will be sparser than that of ZTF \rev{\citep{scoc_report}}, especially when considering sky coverage in a single band, the consecutive observations \rev{over five to six different bands} makes it particularly well suited for the feature extraction approach based on simple parametric multicolor fits like \textsc{Rainbow}. Theoretically, we may expect to get a reasonable estimation of both transient rise time and temperature given just a single point in every filter, separated by a couple of days each -- something that is essentially impossible considering a per-band fitting approach. Typical durations of the rising phase of TDEs, along with immediate availability of \rev{previous alert data for an entire year and} forced photometry at the transient position for at least 30 days prior to the first detection directly in the alert packets will allow us to acquire enough data points for the fit. Thus, we may expect that even without additional support from the local database we will have enough data points for an accurate estimation of the transient physical parameters, and thus will be able to perform its rapid classification.

Given Rubin’s enhanced detection capabilities and the expected LSST alert rate of approximately 10 million per night, the number of candidates reported by the module is expected to increase substantially. To maintain a manageable number of candidates per night for manual vetting, the module may need to be adapted to operate in a more restrictive regime, likely reducing the number of false positives at the cost of a lower completeness.

To train the classifier for Rubin data, it may initially be necessary to rely on simulated datasets, such as those provided by ELAsTiCC \citep{narayan2023extended}. As the survey progresses and a sufficient sample of TDEs is identified by Rubin, the model can be retrained using real observational data. This retraining could also benefit from active learning strategies \citep[e.g.][]{leoni2022fink, moller2025} to efficiently incorporate new labeled examples.

\section{Conclusions}
\label{sec:conclusions}

In this work, we present the development of an early TDE identification module for the \fink\ broker using optical photometry data from ZTF.
The pipeline involves several filtering steps to discard irrelevant alerts and the application of the \textsc{Rainbow} multi-band fitting framework to rising lightcurves in order to extract a set of physically meaningful features.
These features are then used as input to a machine learning classifier trained to distinguish TDEs from other types of rising transients, such as SNe, prior to their peak brightness.

The training dataset consists of 8865 entries that pass all preceding cuts, among which only 42 are labeled as TDEs.
Two variants of the classifier are employed: \rev{the ``broad'' classifier that only uses} three parameters extracted from the fit, and \rev{the ``nuclear'' classifier, which also includes} 
an additional contextual feature, the angular distance to the closest object in the ZTF reference catalog, \texttt{distnr}. 
This value generally improves the model's performance as it helps distinguishing nuclear transients like most TDEs from off-nuclear events such as SNe.
However, because this quantity might not always indicate the distance to the real host and may limit the possibility of identifying off-nuclear TDEs, both models are run in parallel at the current stage.

The \rev{``nuclear''} classifier achieves a recall of 76\% and a precision of 12\%. This performance reflects a design choice: prioritizing completeness (i.e., high recall) over purity to ensure that we capture a broad set of candidates, even at the cost of higher false positives. As a result, manual vetting by experts remains essential to filter down to the most promising sources for multiwavelength follow-up, such as spectroscopy and X-ray observations.
The module is executed in \fink\ after each night of ZTF observations, reporting a list of candidates (currently $1-2$ per night) to dedicated Slack and Telegram channels for visual inspection.

Importantly, the classification is based exclusively on the rising part of a candidate's lightcurve, a design choice whose goal is to enable timely follow-up for the most promising candidates. This goal was successfully achieved, since at least half of the known TDEs identified by our module were flagged before reaching 52\% of their rising phase. 
Such early identification provides ample time to coordinate monitoring of the remaining rise and further follow-up observations around peak brightness.
Increasing the performance in this early identification capability will be one of our goals for future development. 

Throughout the module's development, several interesting transients were identified among the reported candidates when testing on archival ZTF data. While they are likely not all TDEs, many are scientifically interesting worthy of investigation. Section~\ref{sec:results} briefly presents four of these sources, while a more detailed description of these and other transients are presented in the companion article, Quintin \textit{et al.} - in prep.

Despite the considerable effort devoted to the study of TDEs, we still have a long way to go until a fully satisfactorily description of their physical nature can be reached. The advent of large scale sky surveys, like ZTF and soon LSST, enable statistical characterization of such events. This information can be used to fill the gaps of theoretical knowledge with observational data and serve as a starting point for further development. Nevertheless, real time coordination between the many players involved in the data taking and analysis process is crucial to ensure optimized science output from such rich data sets. This work showed one possible approach to ensure optimal exploitation of photometric surveys to enable TDE science. Subsequent work will focus on analysis of a larger sample of candidates\rev{, refinement of the preprocessing steps} and modifications for operation under Rubin. 


\begin{acknowledgements}
We thank Hui Yang for helpful comments on the draft. EQ acknowledges support from the European Space Agency, through the Internal Research Fellowship programme. EQ acknowledges funding from the European
Union’s Horizon 2020 research and innovation programme under grant agreement number 101004168, the XMM2ATHENA project. MVP contribution was carried out under the state assignment of Lomonosov Moscow State University. This work was developed within the \fink\ community and made use of the \fink\ community broker resources. \fink\ is supported by LSST-France and CNRS/IN2P3. 
AM is supported by the Australian Research Council Discovery Early Research Award (DE230100055). Parts of this research were conducted by the Australian Research Council Centre of Excellence for Gravitational Wave Discovery (OzGrav), through project number CE230100016.
This research has made use of the SIMBAD database, operated at CDS, Strasbourg, France.
This work was co-funded by the European Union and supported by the Czech Ministry of Education, Youth and Sports \citep[Project No. CZ.02.01.01/00/22\_008/0004632 -- FORTE,][]{xmm2athena}.
\end{acknowledgements}

%
\bibliographystyle{aa} 
\bibliography{ref} 

\begin{thebibliography}{75}
\expandafter\ifx\csname natexlab\endcsname\relax\def\natexlab#1{#1}\fi

\bibitem[{Aleo {et~al.}(2024)Aleo, Engel, Narayan, Angus, Malanchev, Auchettl, Baldassare, Berres, de~Boer, Boyd, {et~al.}}]{aleo2024anomaly}
Aleo, P., Engel, A., Narayan, G., {et~al.} 2024, The Astrophysical Journal, 974, 172

\bibitem[{Almeida {et~al.}(2023)Almeida, Anderson, Argudo-Fern{\'a}ndez, Badenes, Barger, Barrera-Ballesteros, Bender, Benitez, Besser, Bird, {et~al.}}]{almeida2023eighteenth}
Almeida, A., Anderson, S.~F., Argudo-Fern{\'a}ndez, M., {et~al.} 2023, The Astrophysical Journal Supplement Series, 267, 44

\bibitem[{Bade {et~al.}(1996)Bade, Komossa, \& Dahlem}]{bade_detection_1996}
Bade, N., Komossa, S., \& Dahlem, M. 1996, Astronomy and Astrophysics, 309, L35, aDS Bibcode: 1996A\&A...309L..35B

\bibitem[{{Ban} {et~al.}(2025){Ban}, {Voloshyn}, {Adomavicien{\.{e}}}, {Bachelet}, {Bozza}, {Brincat}, {Bruni}, {Burgaz}, {Carrasco}, {Cassan}, {C̆epas}, {Cusano}, {Dennefeld}, {Dominik}, {Dubois}, {Figuera Jaimes}, {Fukui}, {Galdies}, {Garofalo}, {Hundertmark}, {Ilyin}, {Kruszy{\'n}ska}, {Kulijanishvili}, {Kvernadze}, {Logie}, {Maskoli{\={u}}nas}, {Miko{\l}ajczyk}, {Mr{\'o}z}, {Narita}, {Paks̆tien{\.{e}}}, {Peloton}, {Poleski}, {Qvam}, {Rau}, {Rota}, {Rybicki}, {Street}, {Tsapras}, {Vanaverbeke}, {Wambsganss}, {Wyrzykowski}, {Zdanavic̆ius}, {{\.Z}ejmo}, {Zieli{\'n}ski}, \& {Zola}}]{ban2025}
{Ban}, M., {Voloshyn}, P., {Adomavicien{\.{e}}}, R., {et~al.} 2025, \aap, 697, A57

\bibitem[{Bellm(2022)}]{bellm_zwicky_2014}
Bellm, E. 2022, in The Third Hot-wiring the Transient Universe Workshop, 27--33, conference Name: The Third Hot-wiring the Transient Universe Workshop Pages: 27-33 {ADS} Bibcode: 2014htu..conf...27B

\bibitem[{{Bellm} {et~al.}(2019){Bellm}, {Kulkarni}, {Graham}, {Dekany}, {Smith}, {Riddle}, {Masci}, {Helou}, {Prince}, {Adams}, {Barbarino}, {Barlow}, {Bauer}, {Beck}, {Belicki}, {Biswas}, {Blagorodnova}, {Bodewits}, {Bolin}, {Brinnel}, {Brooke}, {Bue}, {Bulla}, {Burruss}, {Cenko}, {Chang}, {Connolly}, {Coughlin}, {Cromer}, {Cunningham}, {De}, {Delacroix}, {Desai}, {Duev}, {Eadie}, {Farnham}, {Feeney}, {Feindt}, {Flynn}, {Franckowiak}, {Frederick}, {Fremling}, {Gal-Yam}, {Gezari}, {Giomi}, {Goldstein}, {Golkhou}, {Goobar}, {Groom}, {Hacopians}, {Hale}, {Henning}, {Ho}, {Hover}, {Howell}, {Hung}, {Huppenkothen}, {Imel}, {Ip}, {Ivezi{\'c}}, {Jackson}, {Jones}, {Juric}, {Kasliwal}, {Kaspi}, {Kaye}, {Kelley}, {Kowalski}, {Kramer}, {Kupfer}, {Landry}, {Laher}, {Lee}, {Lin}, {Lin}, {Lunnan}, {Giomi}, {Mahabal}, {Mao}, {Miller}, {Monkewitz}, {Murphy}, {Ngeow}, {Nordin}, {Nugent}, {Ofek}, {Patterson}, {Penprase}, {Porter}, {Rauch}, {Rebbapragada}, {Reiley}, {Rigault}, {Rodriguez}, {van Roestel}, {Rusholme}, {van
  Santen}, {Schulze}, {Shupe}, {Singer}, {Soumagnac}, {Stein}, {Surace}, {Sollerman}, {Szkody}, {Taddia}, {Terek}, {Van Sistine}, {van Velzen}, {Vestrand}, {Walters}, {Ward}, {Ye}, {Yu}, {Yan}, \& {Zolkower}}]{bellm2019ztf}
{Bellm}, E.~C., {Kulkarni}, S.~R., {Graham}, M.~J., {et~al.} 2019, \pasp, 131, 018002

\bibitem[{Biswas {et~al.}(2023)Biswas, Ishida, Peloton, M{\"o}ller, Pruzhinskaya, de~Souza, \& Muthukrishna}]{biswas2023enabling}
Biswas, B., Ishida, E., Peloton, J., {et~al.} 2023, Astronomy \& Astrophysics, 677, A77

\bibitem[{Bloom {et~al.}(2011)Bloom, Giannios, Metzger, Cenko, Perley, Butler, Tanvir, Levan, O'Brien, Strubbe, De~Colle, Ramirez-Ruiz, Lee, Nayakshin, Quataert, King, Cucchiara, Guillochon, Bower, Fruchter, Morgan, \& van~der Horst}]{bloom_possible_2011}
Bloom, J.~S., Giannios, D., Metzger, B.~D., {et~al.} 2011, Science, 333, 203, aDS Bibcode: 2011Sci...333..203B

\bibitem[{{Chawla} {et~al.}(2011){Chawla}, {Bowyer}, {Hall}, \& {Kegelmeyer}}]{smote}
{Chawla}, N.~V., {Bowyer}, K.~W., {Hall}, L.~O., \& {Kegelmeyer}, W.~P. 2011, arXiv e-prints, arXiv:1106.1813

\bibitem[{Chen \& Guestrin(2016)}]{xgboost}
Chen, T. \& Guestrin, C. 2016, in Proceedings of the 22nd ACM SIGKDD International Conference on Knowledge Discovery and Data Mining, KDD '16 (New York, NY, USA: Association for Computing Machinery), 785–794

\bibitem[{Dai {et~al.}(2018)Dai, McKinney, Roth, Ramirez-Ruiz, \& Miller}]{dai2018unified}
Dai, L., McKinney, J.~C., Roth, N., Ramirez-Ruiz, E., \& Miller, M.~C. 2018, The Astrophysical Journal Letters, 859, L20

\bibitem[{Drake {et~al.}(2009)Drake, Djorgovski, Mahabal, Beshore, Larson, Graham, Williams, Christensen, Catelan, Boattini, {et~al.}}]{drake2009first}
Drake, A., Djorgovski, S., Mahabal, A., {et~al.} 2009, The Astrophysical Journal, 696, 870

\bibitem[{Duncan(2022)}]{duncan_all-purpose_2022}
Duncan, K.~J. 2022, Monthly Notices of the Royal Astronomical Society, 512, 3662

\bibitem[{{F{\"o}rster} {et~al.}(2021){F{\"o}rster}, {Cabrera-Vives}, {Castillo-Navarrete}, {Est{\'e}vez}, {S{\'a}nchez-S{\'a}ez}, {Arredondo}, {Bauer}, {Carrasco-Davis}, {Catelan}, {Elorrieta}, {Eyheramendy}, {Huijse}, {Pignata}, {Reyes}, {Reyes}, {Rodr{\'\i}guez-Mancini}, {Ruz-Mieres}, {Valenzuela}, {{\'A}lvarez-Maldonado}, {Astorga}, {Borissova}, {Clocchiatti}, {De Cicco}, {Donoso-Oliva}, {Hern{\'a}ndez-Garc{\'\i}a}, {Graham}, {Jord{\'a}n}, {Kurtev}, {Mahabal}, {Maureira}, {Mu{\~n}oz-Arancibia}, {Molina-Ferreiro}, {Moya}, {Palma}, {P{\'e}rez-Carrasco}, {Protopapas}, {Romero}, {Sabatini-Gacitua}, {S{\'a}nchez}, {San Mart{\'\i}n}, {Sep{\'u}lveda-Cobo}, {Vera}, \& {Vergara}}]{Forster:2020}
{F{\"o}rster}, F., {Cabrera-Vives}, G., {Castillo-Navarrete}, E., {et~al.} 2021, \aj, 161, 242

\bibitem[{{Fraga} {et~al.}(2024){Fraga}, {Bom}, {Santos}, {Russeil}, {Leoni}, {Peloton}, {Ishida}, {M{\"o}ller}, \& {Blondin}}]{fraga2024}
{Fraga}, B.~M.~O., {Bom}, C.~R., {Santos}, A., {et~al.} 2024, \aap, 692, A208

\bibitem[{{Gaia Collaboration}(2023)}]{gaia_collaboration_gaia_2023}
{Gaia Collaboration}. 2023, Astronomy \& Astrophysics, 674, A41

\bibitem[{Gezari(2021)}]{gezari_tidal_2021}
Gezari, S. 2021, Ann. Rev. of A\&A, 59, 21, {ADS} Bibcode: 2021ARA\&A..59...21G

\bibitem[{{Gomez} {et~al.}(2023){Gomez}, {Villar}, {Berger}, {Gezari}, {van Velzen}, {Nicholl}, {Blanchard}, \& {Alexander}}]{Gomez2023}
{Gomez}, S., {Villar}, V.~A., {Berger}, E., {et~al.} 2023, \apj, 949, 113

\bibitem[{Gordon(2024)}]{dustmaps}
Gordon, K.~D. 2024, Journal of Open Source Software, 9, 7023

\bibitem[{{Gordon} {et~al.}(2023){Gordon}, {Clayton}, {Decleir}, {Fitzpatrick}, {Massa}, {Misselt}, \& {Tollerud}}]{g23_extinction}
{Gordon}, K.~D., {Clayton}, G.~C., {Decleir}, M., {et~al.} 2023, \apj, 950, 86

\bibitem[{{Guillochon} \& {Ramirez-Ruiz}(2013)}]{Guillochon_Hydrodynamical_2013}
{Guillochon}, J. \& {Ramirez-Ruiz}, E. 2013, \apj, 767, 25

\bibitem[{{Guolo} {et~al.}(2024){Guolo}, {Gezari}, {Yao}, {van Velzen}, {Hammerstein}, {Cenko}, \& {Tokayer}}]{guolo_xraytdes}
{Guolo}, M., {Gezari}, S., {Yao}, Y., {et~al.} 2024, \apj, 966, 160

\bibitem[{Hammerstein {et~al.}(2022)Hammerstein, van Velzen, Gezari, Cenko, Yao, Ward, Frederick, Villanueva, Somalwar, Graham, {et~al.}}]{hammerstein2022final}
Hammerstein, E., van Velzen, S., Gezari, S., {et~al.} 2022, The Astrophysical Journal, 942, 9

\bibitem[{Hung {et~al.}(2017)Hung, Gezari, Blagorodnova, Roth, Cenko, Kulkarni, Horesh, Arcavi, McCully, Yan, {et~al.}}]{hung2017revisiting}
Hung, T., Gezari, S., Blagorodnova, N., {et~al.} 2017, The Astrophysical Journal, 842, 29

\bibitem[{Ivezi{\'c} {et~al.}(2019)Ivezi{\'c}, Kahn, Tyson, Abel, Acosta, Allsman, Alonso, AlSayyad, Anderson, Andrew, {et~al.}}]{ivezic2019lsst}
Ivezi{\'c}, {\v{Z}}., Kahn, S.~M., Tyson, J.~A., {et~al.} 2019, The Astrophysical Journal, 873, 111

\bibitem[{{Kessler} {et~al.}(2009){Kessler}, {Bernstein}, {Cinabro}, {Dilday}, {Frieman}, {Jha}, {Kuhlmann}, {Miknaitis}, {Sako}, {Taylor}, \& {Vanderplas}}]{snana}
{Kessler}, R., {Bernstein}, J.~P., {Cinabro}, D., {et~al.} 2009, \pasp, 121, 1028

\bibitem[{{Langis} {et~al.}(2025){Langis}, {Liodakis}, {Koljonen}, {Paggi}, {Globus}, {Wyrzykowski}, {Miko{\l}ajczyk}, {Kotysz}, {Zieli{\'n}ski}, {Ihanec}, {Ding}, {Morshed}, \& {Torres}}]{langis_tdecatalog_2025}
{Langis}, D.~A., {Liodakis}, I., {Koljonen}, K.~I.~I., {et~al.} 2025, arXiv e-prints, arXiv:2506.05476

\bibitem[{{Le Montagner} {et~al.}(2023){Le Montagner}, {Peloton}, {Carry}, {Desmars}, {Hestroffer}, {Mendez}, {Perlbarg}, \& {Thuillot}}]{lemontagner2023}
{Le Montagner}, R., {Peloton}, J., {Carry}, B., {et~al.} 2023, \aap, 680, A17

\bibitem[{Leoni {et~al.}(2022)Leoni, Ishida, Peloton, \& M{\"o}ller}]{leoni2022fink}
Leoni, M., Ishida, E.~E., Peloton, J., \& M{\"o}ller, A. 2022, Astronomy \& Astrophysics, 663, A13

\bibitem[{Lin {et~al.}(2024)Lin, Jiang, Wang, Kong, Li, He, Wang, Zhu, Li, Jiang, {et~al.}}]{lin2024unluckiest}
Lin, Z., Jiang, N., Wang, T., {et~al.} 2024, The Astrophysical Journal Letters, 971, L26

\bibitem[{Liu {et~al.}(2023)Liu, Malyali, Krumpe, Homan, Goodwin, Grotova, Kawka, Rau, Merloni, Anderson, {et~al.}}]{liu2023deciphering}
Liu, Z., Malyali, A., Krumpe, M., {et~al.} 2023, Astronomy \& Astrophysics, 669, A75

\bibitem[{Liu {et~al.}(2024)Liu, Ryu, Goodwin, Rau, Homan, Krumpe, Merloni, Grotova, Anderson, Malyali, {et~al.}}]{liu2024rapid}
Liu, Z., Ryu, T., Goodwin, A., {et~al.} 2024, Astronomy \& Astrophysics, 683, L13

\bibitem[{Llamas~Lanza {et~al.}(2024)Llamas~Lanza, Quintin, Russeil, Ishida, Peloton, Karpov, \& Pruzhinskaya}]{llamaslanza2024_pTDE}
Llamas~Lanza, M., Quintin, E., Russeil, E., {et~al.} 2024, Transient Name Server AstroNote, 178, 1

\bibitem[{Mahabal {et~al.}(2019)Mahabal, Rebbapragada, Walters, Masci, Blagorodnova, van Roestel, Ye, Biswas, Burdge, Chang, {et~al.}}]{mahabal2019machine}
Mahabal, A., Rebbapragada, U., Walters, R., {et~al.} 2019, Publications of the Astronomical Society of the Pacific, 131, 038002

\bibitem[{{Malanchev} {et~al.}(2023){Malanchev}, {Kornilov}, {Pruzhinskaya}, {Ishida}, {Aleo}, {Korolev}, {Lavrukhina}, {Russeil}, {Sreejith}, {Volnova}, {Voloshina}, \& {Krone-Martins}}]{Malanchev2023}
{Malanchev}, K., {Kornilov}, M.~V., {Pruzhinskaya}, M.~V., {et~al.} 2023, \pasp, 135, 024503

\bibitem[{Malyali {et~al.}(2023)Malyali, Liu, Rau, Grotova, Merloni, Goodwin, Anderson, Miller-Jones, Kawka, Arcodia, {et~al.}}]{malyali2023rebrightening}
Malyali, A., Liu, Z., Rau, A., {et~al.} 2023, Monthly Notices of the Royal Astronomical Society, 520, 3549

\bibitem[{{Masson} \& {Bregeon}(2024)}]{masson2024}
{Masson}, M. \& {Bregeon}, J. 2024, arXiv e-prints, arXiv:2412.05061

\bibitem[{{Masterson} {et~al.}(2024){Masterson}, {De}, {Panagiotou}, {Kara}, {Arcavi}, {Eilers}, {Frostig}, {Gezari}, {Grotova}, {Liu}, {Malyali}, {Meisner}, {Merloni}, {Newsome}, {Rau}, {Simcoe}, \& {van Velzen}}]{masterson_infrared_2024}
{Masterson}, M., {De}, K., {Panagiotou}, C., {et~al.} 2024, \apj, 961, 211

\bibitem[{{Matheson} {et~al.}(2021){Matheson}, {Stubens}, {Wolf}, {Lee}, {Narayan}, {Saha}, {Scott}, {Soraisam}, {Bolton}, {Hauger}, {Silva}, {Kececioglu}, {Scheidegger}, {Snodgrass}, {Aleo}, {Evans-Jacquez}, {Singh}, {Wang}, {Yang}, \& {Zhao}}]{antares}
{Matheson}, T., {Stubens}, C., {Wolf}, N., {et~al.} 2021, AJ, 161, 107

\bibitem[{Milán~Veres {et~al.}(2024)Milán~Veres, Franckowiak, van Velzen, Adebahr, Taziaux, Necker, Stein, Kier, Mueller, Bomans, Jordana-Mitjans, Kowalski, Hammerstein, Marci-Boehncke, Reusch, Garrappa, Rose, \& Kashyap~Das}]{milan_veres_back_2024}
Milán~Veres, P., Franckowiak, A., van Velzen, S., {et~al.} 2024, arXiv e-prints, arXiv:2408.17419

\bibitem[{M{\"o}ller \& de~Boissi{\`e}re(2020)}]{moller2020supernnova}
M{\"o}ller, A. \& de~Boissi{\`e}re, T. 2020, Monthly Notices of the Royal Astronomical Society, 491, 4277

\bibitem[{{M{\"o}ller} {et~al.}(2025){M{\"o}ller}, {Ishida}, {Peloton}, {Vidal Vel{\'a}zquez}, {Soon}, {Martin}, {Cluver}, {Leoni}, \& {Taylor}}]{moller2025}
{M{\"o}ller}, A., {Ishida}, E., {Peloton}, J., {et~al.} 2025, \pasa, 42, e057

\bibitem[{M{\"o}ller {et~al.}(2021)M{\"o}ller, Peloton, Ishida, {et~al.}}]{moller2021fink}
M{\"o}ller, A., Peloton, J., Ishida, E., {et~al.} 2021, Monthly Notices of the Royal Astronomical Society, 501, 3272

\bibitem[{Narayan \& Team(2023)}]{narayan2023extended}
Narayan, G. \& Team, E. 2023, in American Astronomical Society Meeting Abstracts, Vol. 241, American Astronomical Society Meeting Abstracts, 117--01

\bibitem[{{Nordin} {et~al.}(2019){Nordin}, {Brinnel}, {van Santen}, {Bulla}, {Feindt}, {Franckowiak}, {Fremling}, {Gal-Yam}, {Giomi}, {Kowalski}, {Mahabal}, {Miranda}, {Rauch}, {Reusch}, {Rigault}, {Schulze}, {Sollerman}, {Stein}, {Yaron}, {van Velzen}, \& {Ward}}]{ampel}
{Nordin}, J., {Brinnel}, V., {van Santen}, J., {et~al.} 2019, \aap, 631, A147

\bibitem[{Pavez-Herrera {et~al.}(2025)Pavez-Herrera, S{\'a}nchez-S{\'a}ez, Hern{\'a}ndez-Garc{\'\i}a, Bauer, F{\"o}rster, Catelan, Arancibia, Ricci11, Reyes-Jainaga13, Bayo, {et~al.}}]{alerce_tde}
Pavez-Herrera, M., S{\'a}nchez-S{\'a}ez, P., Hern{\'a}ndez-Garc{\'\i}a, L., {et~al.} 2025, arXiv e-prints, arXiv:2503.19698

\bibitem[{Payne {et~al.}(2023)Payne, Auchettl, Shappee, Kochanek, Boyd, Holoien, Fausnaugh, Ashall, Hinkle, Vallely, {et~al.}}]{payne2023chandra}
Payne, A.~V., Auchettl, K., Shappee, B.~J., {et~al.} 2023, The Astrophysical Journal, 951, 134

\bibitem[{Payne {et~al.}(2021)Payne, Shappee, Hinkle, Vallely, Kochanek, Holoien, Auchettl, Stanek, Thompson, Neustadt, {et~al.}}]{payne2021asassn}
Payne, A.~V., Shappee, B.~J., Hinkle, J.~T., {et~al.} 2021, The Astrophysical Journal, 910, 125

\bibitem[{Perley {et~al.}(2023)Perley, Lunnan, Wise, Gkini, Brennan, Pessi, Schulze, Kangas, Sollerman, Yan, {et~al.}}]{perley2023ztf}
Perley, D., Lunnan, R., Wise, J., {et~al.} 2023, Transient Name Server AstroNote, 26, 1

\bibitem[{{Pessi} {et~al.}(2024){Pessi}, {Durgesh}, {Nakazono}, {Hayes}, {Oliveira}, {Ishida}, {Moitinho}, {Krone-Martins}, {Moews}, {de Souza}, {Beck}, {Kuhn}, {Nowak}, \& {Vaughan}}]{pessi2024}
{Pessi}, P.~J., {Durgesh}, R., {Nakazono}, L., {et~al.} 2024, \aap, 691, A181

\bibitem[{Rees(1988)}]{1988Natur.333..523R}
Rees, M.~J. 1988, 333, 523, {ADS} Bibcode: 1988Natur.333..523R

\bibitem[{Russeil {et~al.}(2024)Russeil, {Malanchev, K. L.}, {Aleo, P. D.}, {Ishida, E. E. O.}, {Pruzhinskaya, M. V.}, {Gangler, E.}, {Lavrukhina, A. D.}, {Volnova, A. A.}, {Voloshina, A.}, {Semenikhin, T.}, {Sreejith, S.}, {Kornilov, M. V.}, \& {Korolev, V. S.}}]{russeil2024}
Russeil, E., {Malanchev, K. L.}, {Aleo, P. D.}, {et~al.} 2024, A\&A, 683, A251

\bibitem[{{Russeil} {et~al.}(2024){Russeil}, {Quintin}, {Llamas Lanza}, {Ishida}, {Peloton}, {Karpov}, {Pruzhinskaya}, {S}, {A}, \& {A}}]{tde_2020ukj}
{Russeil}, E., {Quintin}, E., {Llamas Lanza}, M., {et~al.} 2024, Transient Name Server Classification Report, 2024-5006, 1

\bibitem[{S{\'a}nchez-S{\'a}ez {et~al.}(2024)S{\'a}nchez-S{\'a}ez, Hern{\'a}ndez-Garc{\'\i}a, Bernal, Bayo, Rivera, Bauer, Ricci, Merloni, Graham, Cartier, {et~al.}}]{sanchez2024sdss1335}
S{\'a}nchez-S{\'a}ez, P., Hern{\'a}ndez-Garc{\'\i}a, L., Bernal, S., {et~al.} 2024, Astronomy \& Astrophysics, 688, A157

\bibitem[{{S{\'a}nchez-S{\'a}ez} {et~al.}(2021){S{\'a}nchez-S{\'a}ez}, {Reyes}, {Valenzuela}, {F{\"o}rster}, {Eyheramendy}, {Elorrieta}, {Bauer}, {Cabrera-Vives}, {Est{\'e}vez}, {Catelan}, {Pignata}, {Huijse}, {De Cicco}, {Ar{\'e}valo}, {Carrasco-Davis}, {Abril}, {Kurtev}, {Borissova}, {Arredondo}, {Castillo-Navarrete}, {Rodriguez}, {Ruz-Mieres}, {Moya}, {Sabatini-Gacit{\'u}a}, {Sep{\'u}lveda-Cobo}, \& {Camacho-I{\~n}iguez}}]{alerce_previous}
{S{\'a}nchez-S{\'a}ez}, P., {Reyes}, I., {Valenzuela}, C., {et~al.} 2021, \aj, 161, 141

\bibitem[{Saxton {et~al.}(2018)Saxton, Perets, \& Baskin}]{saxton2018spectral}
Saxton, C.~J., Perets, H.~B., \& Baskin, A. 2018, Monthly Notices of the Royal Astronomical Society, 474, 3307

\bibitem[{{Schlegel} {et~al.}(1998){Schlegel}, {Finkbeiner}, \& {Davis}}]{sfd_map}
{Schlegel}, D.~J., {Finkbeiner}, D.~P., \& {Davis}, M. 1998, \apj, 500, 525

\bibitem[{SCOC(2025)}]{scoc_report}
SCOC, T. R. O. S. C. O.~C. 2025, Survey Cadence Optimization Committee’s Phase 3 Recommendations, Tech. Rep. PSTN-056

\bibitem[{Shatsky {et~al.}(2020)Shatsky, Belinski, Dodin, Zheltoukhov, Kornilov, Postnov, Potanin, Safonov, Tatarnikov, \& Cherepashchuk}]{shatsky2020caucasian}
Shatsky, N., Belinski, A., Dodin, A., {et~al.} 2020, arXiv preprint arXiv:2010.10850

\bibitem[{{Sheng} {et~al.}(2024){Sheng}, {Nicholl}, {Smith}, {Young}, {Williams}, {Stevance}, {Smartt}, {Srivastav}, \& {Moore}}]{needle}
{Sheng}, X., {Nicholl}, M., {Smith}, K.~W., {et~al.} 2024, \mnras, 531, 2474

\bibitem[{Shlentsova {et~al.}(2024)Shlentsova, Hoof, Dalen, Fraser, Barmentloo, Kachiprath, Kumar, Anderson, Chen, Gromadzki, {et~al.}}]{shlentsova2024epessto}
Shlentsova, A., Hoof, A., Dalen, J., {et~al.} 2024, Transient Name Server AstroNote, 98, 1

\bibitem[{Smith {et~al.}(2019)Smith, Williams, Young, Ibsen, Smartt, Lawrence, Morris, Voutsinas, \& Nicholl}]{lasair}
Smith, K.~W., Williams, R.~D., Young, D.~R., {et~al.} 2019, Research Notes of the {AAS}, 3, 26

\bibitem[{Somalwar {et~al.}(2025{\natexlab{a}})Somalwar, {Ravi}, {Margutti}, {Chornock}, {Natarajan}, {Lu}, {Angus}, {Graham}, {Hammerstein}, {Nathan}, {Nicholl}, {Sharma}, {Stein}, {Verdi}, {Yao}, {Bellm}, {Chen}, {Coughlin}, {Hale}, {Kasliwal}, {Laher}, {Riddle}, \& {Sollerman}}]{offnuclear_AT2024puz}
Somalwar, J.~J., {Ravi}, V., {Margutti}, R., {et~al.} 2025{\natexlab{a}}, arXiv e-prints, arXiv:2505.11597

\bibitem[{Somalwar {et~al.}(2025{\natexlab{b}})Somalwar, Ravi, Yao, Guolo, Graham, Hammerstein, Lu, Nicholl, Sharma, Stein, {et~al.}}]{somalwar2025first}
Somalwar, J.~J., Ravi, V., Yao, Y., {et~al.} 2025{\natexlab{b}}, The Astrophysical Journal, 985, 175

\bibitem[{{Stein} {et~al.}(2023){Stein}, {Mahabal}, {Reusch}, {Graham}, {Kasliwal}, {Kowalski}, {Gezari}, {Hammerstein}, {Nakoneczny}, {Nicholl}, {Sollerman}, {van Velzen}, {Yao}, {Laher}, \& {Rusholme}}]{Stein_tdescore_2023}
{Stein}, R., {Mahabal}, A., {Reusch}, S., {et~al.} 2023, arXiv e-prints, arXiv:2312.00139

\bibitem[{Sun {et~al.}(2025)Sun, Guo, Gu, Li, Chen, González-Buitrago, Wang, Li, Feng, Xiong, Wang, Yuan, Jin, Zhang, Deng, \& Zhang}]{sun_at2021aeuk_2025}
Sun, J., Guo, H., Gu, M., {et~al.} 2025, The Astrophysical Journal, 982, 150

\bibitem[{Sun {et~al.}(2024)Sun, Jiang, Dou, Shu, Zhu, Dong, Buckley, Bradley~Cenko, Fan, Gromadzki, Liu, Wang, Wang, Wang, Wu, Yang, Zhang, Zhang, \& Zhang}]{sun_recurring_2024}
Sun, L., Jiang, N., Dou, L., {et~al.} 2024, Astronomy and Astrophysics, 692, A262

\bibitem[{van Velzen {et~al.}(2020)van Velzen, Holoien, Onori, Hung, \& Arcavi}]{van2020optical}
van Velzen, S., Holoien, T. W.-S., Onori, F., Hung, T., \& Arcavi, I. 2020, Space science reviews, 216, 1

\bibitem[{{Webb} {et~al.}(2023){Webb}, {Carrera}, {Schwope}, {Motch}, {Ballet}, {Watson}, {Page}, {Freyberg}, {Georgantopoulos}, {Coriat}, {Barret}, {Massida}, {Gupta}, {Tranin}, {Quintin}, {Teresa Ceballos}, {Mateos}, {Corral}, {Dominguez}, {Stiele}, {Traulsen}, {Pires}, {Nebot}, {Michel}, {Xavier Pineau}, {Kuuttila}, {Maggi}, {Chakroborty}, {Birchall}, {Kuin}, {Akylas}, {Ruiz}, {Pouliasis}, \& {Georgakakis}}]{xmm2athena}
{Webb}, N.~A., {Carrera}, F.~J., {Schwope}, A., {et~al.} 2023, Astronomische Nachrichten, 344, e20220102

\bibitem[{Wenger {et~al.}(2000)Wenger, Ochsenbein, Egret, Dubois, Bonnarel, Borde, Genova, Jasniewicz, Lalo{\"e}, Lesteven, {et~al.}}]{wenger2000simbad}
Wenger, M., Ochsenbein, F., Egret, D., {et~al.} 2000, Astronomy and Astrophysics Supplement Series, 143, 9

\bibitem[{Wevers {et~al.}(2023)Wevers, Coughlin, Pasham, Guolo, Sun, Wen, Jonker, Zabludoff, Malyali, Arcodia, {et~al.}}]{wevers2023live}
Wevers, T., Coughlin, E., Pasham, D., {et~al.} 2023, The Astrophysical Journal Letters, 942, L33

\bibitem[{Wright {et~al.}(2010)Wright, Eisenhardt, Mainzer, Ressler, Cutri, Jarrett, Kirkpatrick, Padgett, McMillan, Skrutskie, {et~al.}}]{wright2010wide}
Wright, E.~L., Eisenhardt, P.~R., Mainzer, A.~K., {et~al.} 2010, The Astronomical Journal, 140, 1868

\bibitem[{Yao {et~al.}(2025)Yao, Chornock, Ward, Hammerstein, Sfaradi, Margutti, Kelley, Lu, Liu, Wise, {et~al.}}]{yao2025offnuclear_AT2024tvd}
Yao, Y., Chornock, R., Ward, C., {et~al.} 2025, The Astrophysical Journal Letters, 985, L48

\bibitem[{{Yao} {et~al.}(2023){Yao}, {Ravi}, {Gezari}, {van Velzen}, {Lu}, {Schulze}, {Somalwar}, {Kulkarni}, {Hammerstein}, {Nicholl}, {Graham}, {Perley}, {Cenko}, {Stein}, {Ricarte}, {Chadayammuri}, {Quataert}, {Bellm}, {Bloom}, {Dekany}, {Drake}, {Groom}, {Mahabal}, {Prince}, {Riddle}, {Rusholme}, {Sharma}, {Sollerman}, \& {Yan}}]{yao_ztfdemographics_2023}
{Yao}, Y., {Ravi}, V., {Gezari}, S., {et~al.} 2023, \apjl, 955, L6

\bibitem[{{Yuan} {et~al.}(2015){Yuan}, {Zhang}, {Feng}, {Zhang}, {Ling}, {Zhao}, {Deng}, {Qiu}, {Osborne}, {O'Brien}, {Willingale}, {Lapington}, {Fraser}, \& {the Einstein Probe team}}]{einsteinProbe_2015}
{Yuan}, W., {Zhang}, C., {Feng}, H., {et~al.} 2015, arXiv e-prints, arXiv:1506.07735

\end{thebibliography}
%

\appendix
\rev{
\section{Illustration of lightcurve window selection and fit}}

\rev{
This appendix aims to show a visual example of the different lightcurve windows defined in Section~\ref{subsec:prepro} (fitting, historical and pre-historical), together with the corresponding \textsc{Rainbow} fit described in Section~\ref{sec:features}.
To ease the visualization, we selected an object classified as Seyfert 1 exhibiting a rising segment with clear preceding activity.}

\begin{figure}[h!]
\centering
\includegraphics[width=1\columnwidth]{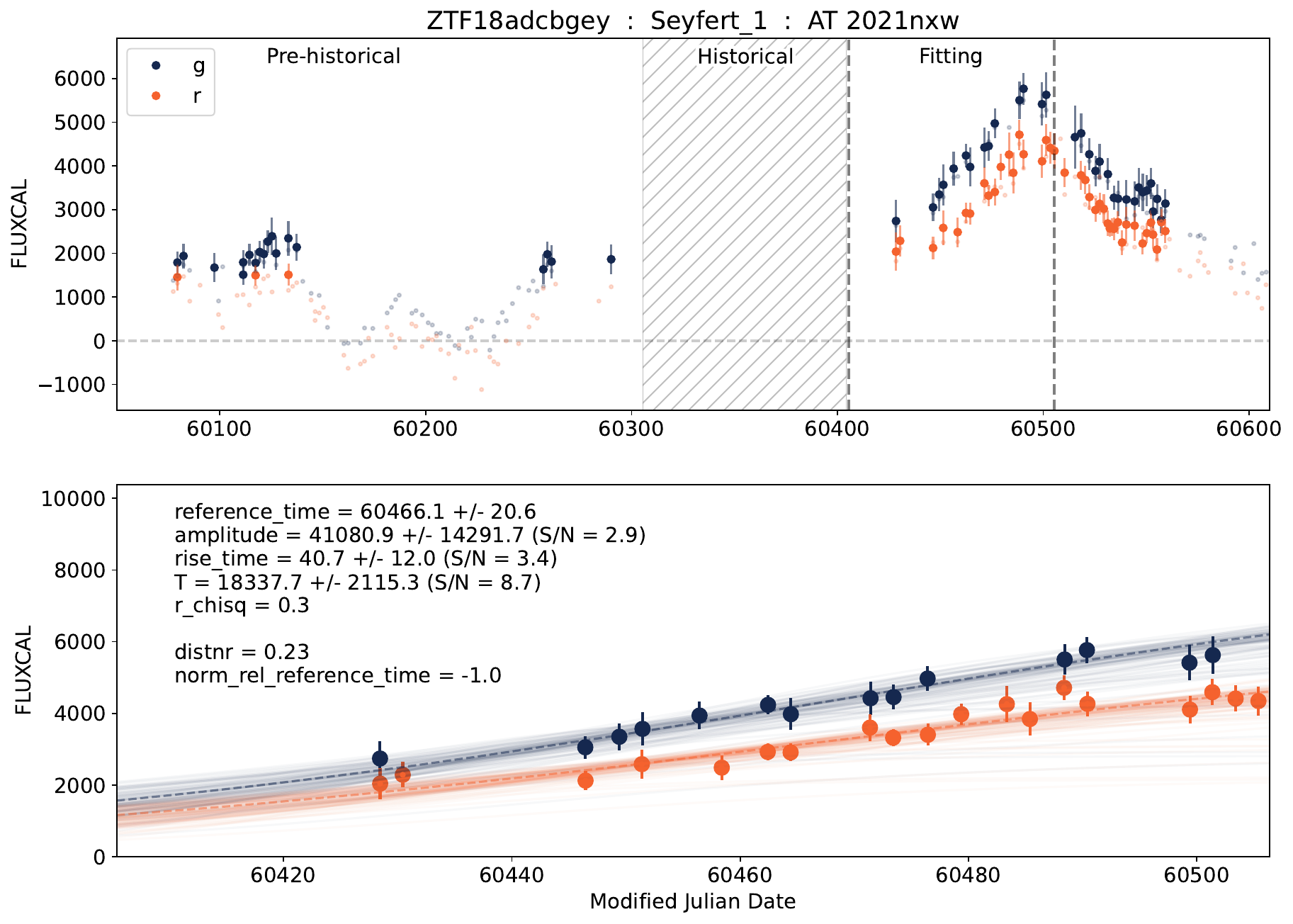}
\caption{Illustration on the window selection and feature extraction processes. 
The upper panel shows the complete lightcurve extracted from the alerts (full circles) and from data-release photometry (translucid dots) with the fitting, historical and pre-historical windows described in Section~\ref{subsec:prepro} indicated for context. 
Lower panel displays the data points within the 100-day fitting window, along with the corresponding \textsc{Rainbow} fit output (Section~\ref{sec:features}). The dashed lines correspond to the best fit parameter values, while the translucid ones show the scatter of models with parameters sampled around the best fit according to the estimated covariances. The in-figure annotation lists the values of the best fit parameters along with their uncertainties, as well as some additional features such as \texttt{distnr}.}
\label{fig:rainbow_example}
\end{figure}

\rev{
The upper panel of Figure~\ref{fig:rainbow_example} shows the full lightcurve with alerts represented by full circles and data-release photometry by translucid dots. 
Black vertical dashed lines mark the 100-day rising interval for fitting, the hatched region marks the immediate historical 100-day window prior to that, used to constrain ongoing continuous activity, and everything prior to that is the pre-historical period used to additionally restrict long-term behavior.
The window selection shown here corresponds to the last alert that passed the criteria presented in Section~\ref{subsec:prepro} (posterior alerts are significantly decaying). Note in order to allow for repeated TDE flares, pre-historical activity is allowed, if there is at most one point with negative flux.}

\rev{The lower panel of Figure~\ref{fig:rainbow_example} shows the data points within the fitting window together with the results of the corresponding \textsc{Rainbow} fit. The best-fit parameter values and corresponding uncertainties are annotated on the figure for reference.}

\end{document}